\documentclass[twocolumn]{aa}

\usepackage{graphicx}
\usepackage{color,txfonts}
\usepackage{microtype}
\usepackage[colorlinks=true,unicode=true]{hyperref}
\hypersetup{citecolor= blue, linkcolor=red, urlcolor=blue}

\newcommand{\Hazel}{{\sc Hazel}}
\newcommand{\Sir}{{\sc Sir}}
\newcommand{\MultiNest}{{\sc MultiNest}}
\newcommand{\PyMultiNest}{{\sc PyMultiNest}}
\newcommand{\Helix}{{\sc HeLIx}}
\newcommand{\kms}{\,km\,s$^{-1}$}

\newcommand{\thetabold}{\mbox{\boldmath$\theta$}}
\newcommand{\Sbold}{\mbox{\boldmath$S$}}

\begin{document} 

   \title{Spectropolarimetric analysis of an active region filament}
   \subtitle{I. Magnetic and dynamical properties from single component inversions}
 
   \author{D\'iaz Baso, C. J.
          \inst{1,2,3}
          \and
          Mart\'{\i}nez Gonz\'alez, M. J.
          \inst{1,2}
          \and
          Asensio Ramos,  A.
          \inst{1,2}
          }

   \institute{Instituto de Astrof\'isica de Canarias, C/V\'{\i}a L\'actea s/n, E-38205 La Laguna, Tenerife, Spain
   \and
   Departamento de Astrof\'{\i}­sica, Universidad de La Laguna, E-38206 La Laguna, Tenerife, Spain
   \and
   Institute for Solar Physics, Dept. of Astronomy, Stockholm University, AlbaNova University Centre, SE-10691 Stockholm Sweden
             }

   \date{Received December 06, 2018; accepted April 12, 2019}
   
   \authorrunning{D\'iaz Baso et al.}
   \titlerunning{Magnetic field of an active region filament}
 
  \abstract
   {}
   {The determination of the magnetic filed vector in solar filaments is possible by interpreting the Hanle and Zeeman effects in suitable chromospheric spectral lines like those of the \ion{He}{i} multiplet at 10830\,\AA. We study the vector magnetic field of an active region filament (NOAA 12087).}
  {Spectropolarimetric data of this active region was acquired with the GRIS instrument at the GREGOR telescope and studied simultaneously in the chromosphere with the \ion{He}{i} 10830\,\AA\ multiplet and in the photosphere \ion{Si}{i} 10827\,\AA\ line. As it is usual from previous studies, only a single component model is used to infer the magnetic properties of the filament. The results are put into a solar context with the help of the Solar Dynamic Observatory images.}
  {Some results clearly point out that a more complex inversion had to be done. Firstly, the Stokes $V$ map of \ion{He}{i} does not show any clear signature of the presence of the filament. Secondly, the local azimuth map follows the same pattern than Stokes $V$ as if the polarity of Stokes $V$ were conditioning the inference to very different magnetic field even with similar linear polarization signals. This indication suggests that the Stokes $V$ could be dominated by the below magnetic field coming from the active region, and not, from the filament itself. Those and more evidences will be analyzed in depth and a more complex inversion will be attempted in the second part of this series.}
  {}


   \maketitle

\section{Introduction}\label{sec:intro}

Solar filaments and prominences are cool chromospheric over-dense plasma structures embedded in the extremely hot and less dense corona. They appear as dark filamentary structures when seen on the disk as absorption in the core of some strong chromospheric lines (such as H$\alpha$ or \ion{Ca}{ii} lines), other weak chromospheric lines such as the \ion{He}{i} multiplets at 10830\,\AA\ and 5876\,\AA\ (D$_3$), and in the extreme ultraviolet continua. They are called prominences when they are seen as diffuse bright clouds at the limb, as they scatter light from the underlying disk. 

Magnetic fields play a fundamental role in the formation, support, and eruption of these structures. The topology of the magnetic field is a key point because the magnetic force can balance the solar gravitational force and support the plasma in places where the fields are upward-curved, forming dips. Even being such a fundamental ingredient, its topology is not well known. Different theoretical models have been developed proposing magnetic configurations with dipped field lines. Magnetic dips can be created by the weight of the prominence plasma \citep{Kippenhahn1957}; or they can exist in the absence of prominence plasma in force-free models as an horizontal flux rope \citep{Aulanier1998,Kuperus1974} or a sheared arcade \citep{Antiochos1994}; or even by a tangled magnetic field \citep{vanB2010} to explain prominences with vertical threads.

While intensity observations of prominence fine structure provide some clues to the magnetic topology, the only method to constrain  theoretical models is the empirical determination of the prominence magnetic field from the inversion of spectro-polarimetric data \citep{Harvey2006}. The inference of prominence magnetic fields is not straightforward since there are only few chromospheric spectral lines with sufficient opacity to sample the chromosphere which, in turn, have such a wide and high formation region that have to be modeled under complex NLTE effects, such as the atomic polarization and the Hanle effect \citep{landi_landolfi04}.

The \ion{He}{i} triplets at 10830\,\AA\ and 5876\,\AA\ are a suitable pair of diagnostics for such studies since they present a good compromise between sensitivity to chromospheric magnetic fields and complexity of the numerical modeling. The \ion{He}{i} 10830\,\AA\ multiplet has several advantages: it can be easily seen in emission in off-limb prominences \citep[e.g.,][]{orozco2014, marian2015} or in absorption in on-disk filaments \citep[e.g.,][]{lin1998, Collados2003, merenda2007, kuckein2009, sasso2011} and it is sensitive to magnetic fields in a wide range from dG to kG thanks to the joint action of atomic level polarization, the Hanle and Zeeman effects. These lines have been used since the 1970s \citep{leroy1977}, and they were consolidated as a very useful diagnostic tool for these type of structures after \cite{TrujilloBueno2002} demonstrated how the polarized light arises from selective absorption and emission processes induced by the solar anisotropic illumination and its dependence on the scattering geometry. Since then, these spectral lines have been increasingly used to infer the magnetic field vector in chromospheric structures.

Filaments can be categorized by their location on the Sun and are generally considered to have very different magnetic structures and formation processes. In one limit we find quiescent filaments (QS), with a lifetime of up to a few months, hundreds of Mm of length and height, and magnetic fields strengths of the order of a few tens of G \citep{TrujilloBueno2002, casini2003, merenda2006, orozco2014, marian2015}. On the other limit, we find active region (AR) filaments, that are formed on active regions or sunspots, they develop at lower heights in the solar atmosphere \citep[$ <10 $ Mm, ][]{Aulanier2003, Lites2005}, have small lifetimes (hours to days), have short lengths (over 10 Mm) and have magnetic fields larger than that of QS filaments. Sometimes the term intermediate filaments is used when no clear classification into QS or AR is possible \citep{mackay2010}.

The linear polarization of the \ion{He}{i} 10830\,\AA\ in QS filaments is dominated by atomic level polarization induced by the anisotropic illuminating radiation field coming from lower layers \citep{TrujilloBueno2002, marian2015}. However, AR filaments have been found under a higher variety of conditions. For example, \cite{kuckein2009} studied the magnetic field vector of an active region filament above a strongly magnetized region. Their signals were dominated by the Zeeman effect without almost any evidence of atomic polarization. They found, with a Zeeman-based Milne-Eddington (ME) inversion code, the highest field strengths measured in filaments to date, with values around 600--700\,G. Their conclusion was that strong transverse magnetic fields were present in AR filaments. After that, \cite{sasso2011, sasso2014} studied a complex flaring active region filament during its phase of increased activity with an ME inversion code \cite[\Helix; ][]{lagg2004} including the Hanle effect, as they detected atomic level polarization signatures in their observed profiles. They found values for the magnetic field strength in the range 100--250\,G, requiring up to five independent atmospheric components coexisting within the same resolution element ($\sim$1.2'') to reproduce their line profiles. Another measurement was reported by \cite{xu2012}, who inverted the same filament as \cite{sasso2011} but before the eruption, during the stable phase. They used again the Zeeman-based ME inversion \cite[\Helix$^+$;][]{lagg2009} because they did not find any clear signatures of atomic level polarization nor the Hanle effect. Their observation displayed a section of the filament above the granulation with weaker inferred magnetic fields ($\sim$150\,G) and a second section above a strongly magnetized area,  which returned much higher magnetic field strengths during the inversion ($\sim$750\,G).

At that time, the non-detection of signatures of the presence of atomic level polarization was quite surprising \citep{kuckein2009,xu2012}. \cite{TB2007} and \cite{CASINI2009} noted soon that, even for such strong fields, AR filaments spectropolarimetric observations should show signatures of scattering polarization and the Hanle effect in normal conditions. To explain this fact, \cite{TB2007} suggested that a reduction of the radiation field anisotropy due to the horizontal radiation inside the filament could reduce the atomic level polarization, while \cite{CASINI2009} suggested that an unresolved magnetic field could also explain the observation.

Taking into account all these measurements, the highest inferred magnetic field values seem to be associated with filaments above highly magnetized areas. Based on this idea, \cite{diaz2016} suggested that filaments could be quite transparent to polarization, with the highest values for the magnetic field being produced by the active chromosphere below, and not from the filament itself. By adopting a two-component model (one slab for the filament and another for the chromosphere below) these authors reproduce the profiles in \cite{kuckein2009} without suppressing the  atomic polarization. The lack of atomic polarization found in \cite{kuckein2009} could be due to a cancellation of the total atomic polarization signal from the two corresponding components. From their modeling, weak fields can be also inferred in AR filaments.

In the light of these results the magnetic field strength inferred in prominences is still a matter of debate. At present, a high quality polarimetric observation is a challenge and the inference of the magnetic field vector in the Hanle regime is subject to several potential ambiguities \citep{orozco2014,marian2015}.

To shed some light and better understand the problems during the inference, we present a first analysis of ground-based spectropolarimetric observations of the \ion{He}{i} 10830 triplet taken in an active region filament. We first describe the observations and then explain the diagnostic technique. We study the observations with state-of-the-art inversion codes including the Hanle effect and the Zeeman effect to retrieve the physics from the polarimetric signals. During this study we show some limitations of using a one component model as it has been common practice.  After that, we use the Bayesian framework to study the Hanle  ambiguities and the uncertainties and the sensitivity of some parameters in the inference.

\begin{figure}
\centering
\includegraphics[width=\linewidth]{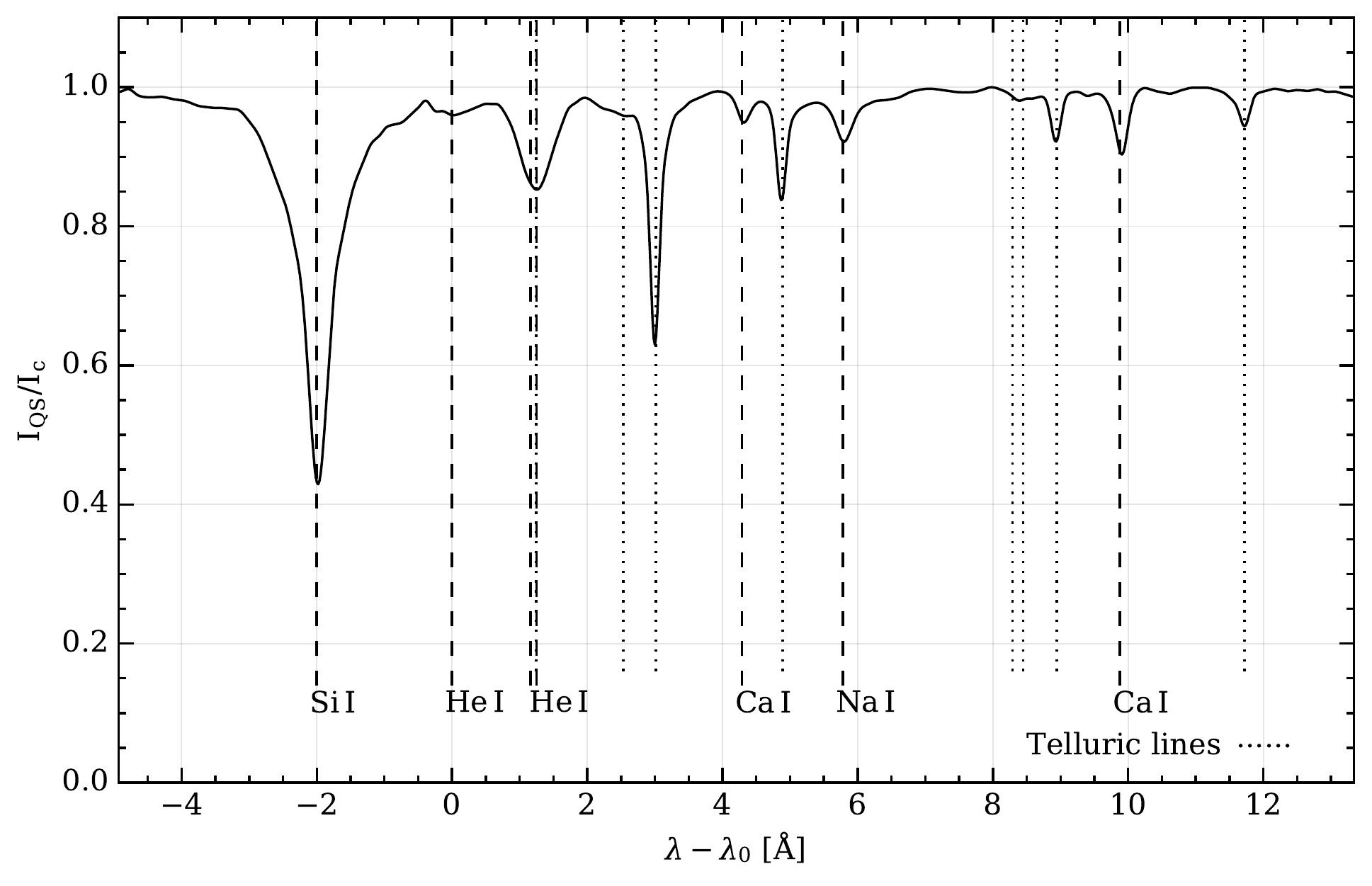}
\caption{This figure displays the average intensity spectrum of the quiet-Sun showing the GRIS spectral range and the mentioned photospheric and chromospheric spectral lines. The reference wavelength ($\lambda_{0}=10829.0911$\,\AA)  
is the central wavelength of the blue component of the \ion{He}{i} 10830\,\AA\ multiplet. The symbol $I_c$ is referring to the continuum intensity.}
\label{fig:RANGO1}
\end{figure}

\section{Observations, context data and filament evolution}\label{sec:observation}

The observations were carried out at the GREGOR telescope \citep{gregor2012} on the 17th of June 2014 using GRIS  \citep[GREGOR Infrared Spectrograph,][]{gris2012}. The slit (0.14\arcsec\ wide and 65\arcsec\ long) was placed over the target, an active region filament close to a sunspot in the active region NOAA 12087, at coordinates +187\arcsec,-327\arcsec, which correspond to an  heliocentric angle of $\theta=23^\circ$ ($\mu=\mathrm{cos}\theta=0.92$).  Six scans of the full sunspot were taken with a scanning step of 0.14\arcsec, which provided us with a map of size 65\arcsec\ $\times$ 27\arcsec. The first scan started at 08:00 UT and the last one finished at 17:30 UT. Each one lasts $\sim$15 minutes. The rotation compensation of the alt-azimuthal mount in GREGOR was not yet available, so we witnessed some image rotation effects, specially in the borders of the slit. By calculating the power spectrum of the continuum intensity in the quiet Sun we estimated the spatial resolution to be about 0.4\arcsec.

\begin{figure}[h]
\centering
\includegraphics[width=\linewidth]{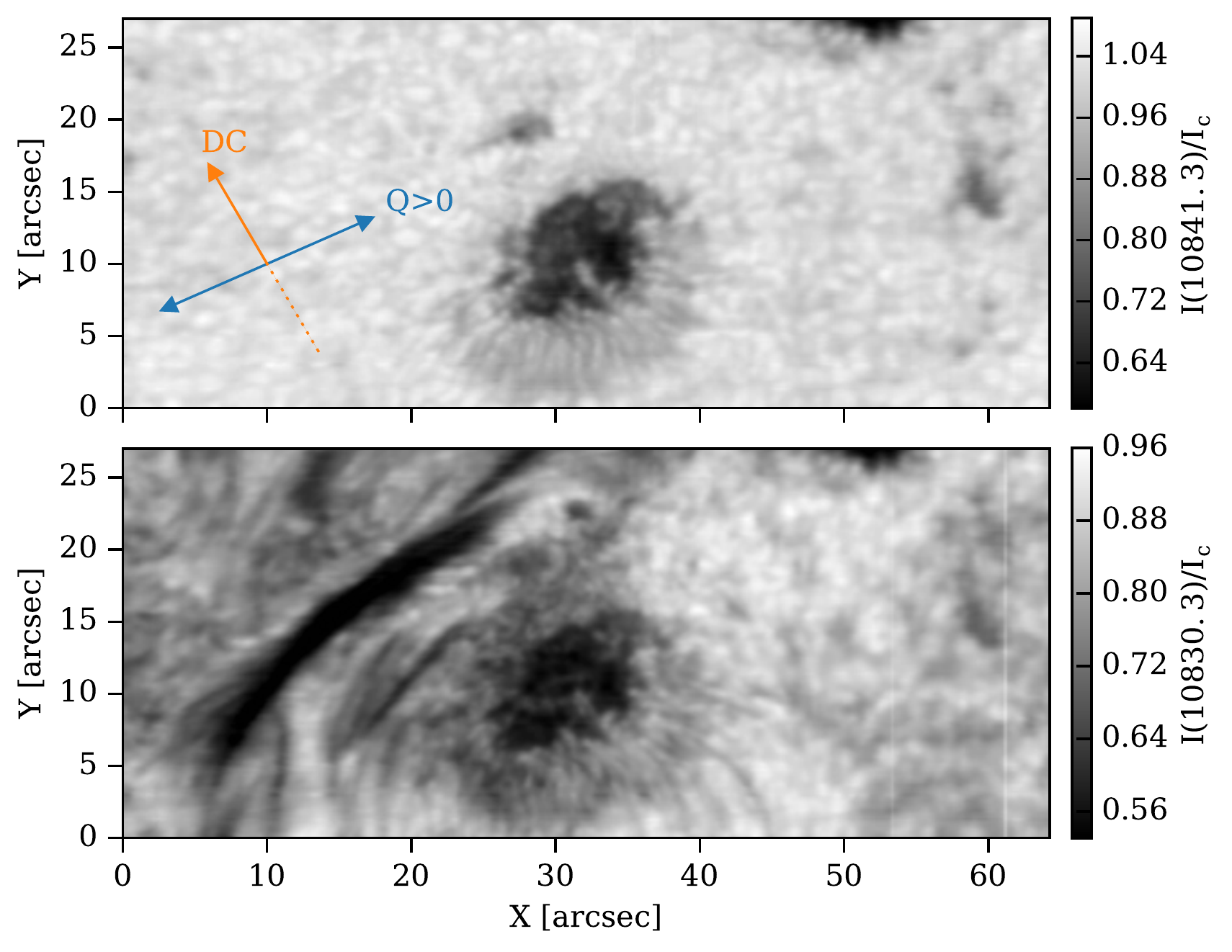}
\caption{Reconstructed maps of the Continuum map (top panel) and the \ion{He}{i} 10830\,\AA\ red core intensity (bottom panel) taken at 9:45 UT. The orange arrow indicate the direction towards the disk center (DC) and the blue one represents the reference of positive  Stokes $Q$.}
\label{fig:MAPA2}
\end{figure}

The observed spectral range spanned from 10824\,\AA\ to 10842\,\AA, with a spectral sampling of 18.1 m\AA/pixel. This spectral region contains the chromospheric \ion{He}{i} triplet, some photospheric lines of \ion{Si}{i}, \ion{Ca}{i} and \ion{Na}{i}, and several telluric lines. The \ion{He}{i} 10830\,\AA\ triplet, originates from the electronic transition 2\,$^3S_1$ $\to$ 2\,$^3P_{2,1,0}$. It comprises a component at 10829.0911\AA\ ($J_u=0$), usually known as the ``blue'' component, and two components at 10830.2501\,\AA\ ($J_u=1$) and at 10830.3397\,\AA\ ($J_u=2$), blended at solar chromospheric temperatures, and usually known as the ``red'' component. The GRIS spectral range with the above-mentioned spectral lines is displayed in Fig.~\ref{fig:RANGO1} together with the average intensity spectrum of the quiet Sun of our observations.

We show one of the reconstructed maps of the observed active region  in Fig.~\ref{fig:MAPA2}, where the x-axis represents the position along the slit and the y-axis is the scanning direction. The upper panel of the figure shows the Stokes $I$ map of the observed region in the continuum at 10841.3\,\AA, and the lower panel shows the same region in the core of the red component of the helium triplet at 10830.3\,\AA, normalized to the continuum intensity of the quiet Sun. The filament is visible in the core of the \ion{He}{i} line with a length of about $\sim$30\arcsec (22\,Mm) and a width of $\sim$3\arcsec (2\,Mm). These images show that the observations were carried out under very good seeing conditions. Both the granulation pattern and chromospheric absorption features with sub-arcsecond fine structure are prominently visible.

\subsection{Reduction process}

The standard data reduction procedure for GRIS \citep{Collados1999,Collados2003R} was applied to all the data sets. This reduction includes bad pixel correction, dark current subtraction, flat-field correction, telescope and polarimeter calibration, demodulation, beam merging and correction of residual fringes. To improve the signal to noise ratio of the data we averaged 2 spectral pixels and 4$\times$4 spatial pixels, yielding a final spectral and spatial sampling of 36.2 m\AA~pix$^{-1}$ and 0.55\arcsec~pix$^{-1}$, respectively, and a noise level in Stokes ($Q$, $U$, $V$) estimated from the standard deviation of the continuum, of the order of $(4, 4, 6)\times 10^{-4}$ in units of the continuum intensity, $I_c$.

The next step is to correct spurious light trends and fringes that persist after the flat-fielding. The continuum intensity trends are corrected by comparison with the Fourier Transform Spectrometer \citep[$I_{\mathrm{FTS}}(\lambda)$ or FTS,][]{neckel1984} atlas. This atlas has a very high signal-to-noise ratio, no spectral stray light, and very accurate continuum level. Following \cite{Allende2004} and \cite{Franz2016}, the $I_{\mathrm{FTS}}(\lambda)$ is modified to be comparable to our spectrum $I_{QS}(\lambda)$, generating a "new" intensity profile $I_{\mathrm{NFTS}}(\lambda)$:
\begin{equation}
I_{\mathrm{NFTS}}(\lambda) = \alpha\langle G(\sigma)*I_{\mathrm{FTS}}(\lambda)\rangle + (1-\alpha)G(\sigma)*I_{\mathrm{FTS}}(\lambda) \ .
\label{eq:calibration}
\end{equation}

In order to take into account the different spectral resolution of our instrument, $I_{\mathrm{FTS}}(\lambda)$ is convolved with a Gaussian $G(\sigma)$ with a standard deviation $\sigma$. Then, we add another term $\langle G(\sigma)*I_{\mathrm{FTS}}(\lambda)\rangle$ which is the average over the same spectral range of the convolved spectrum and it does not depend on the wavelength. This allows us to characterize the amount of non-dispersed spectrum (light that reaches the detector without passing through the grating) by the stray light factor $\alpha$.

\begin{figure}[!ht]
\centering
\includegraphics[width=\linewidth]{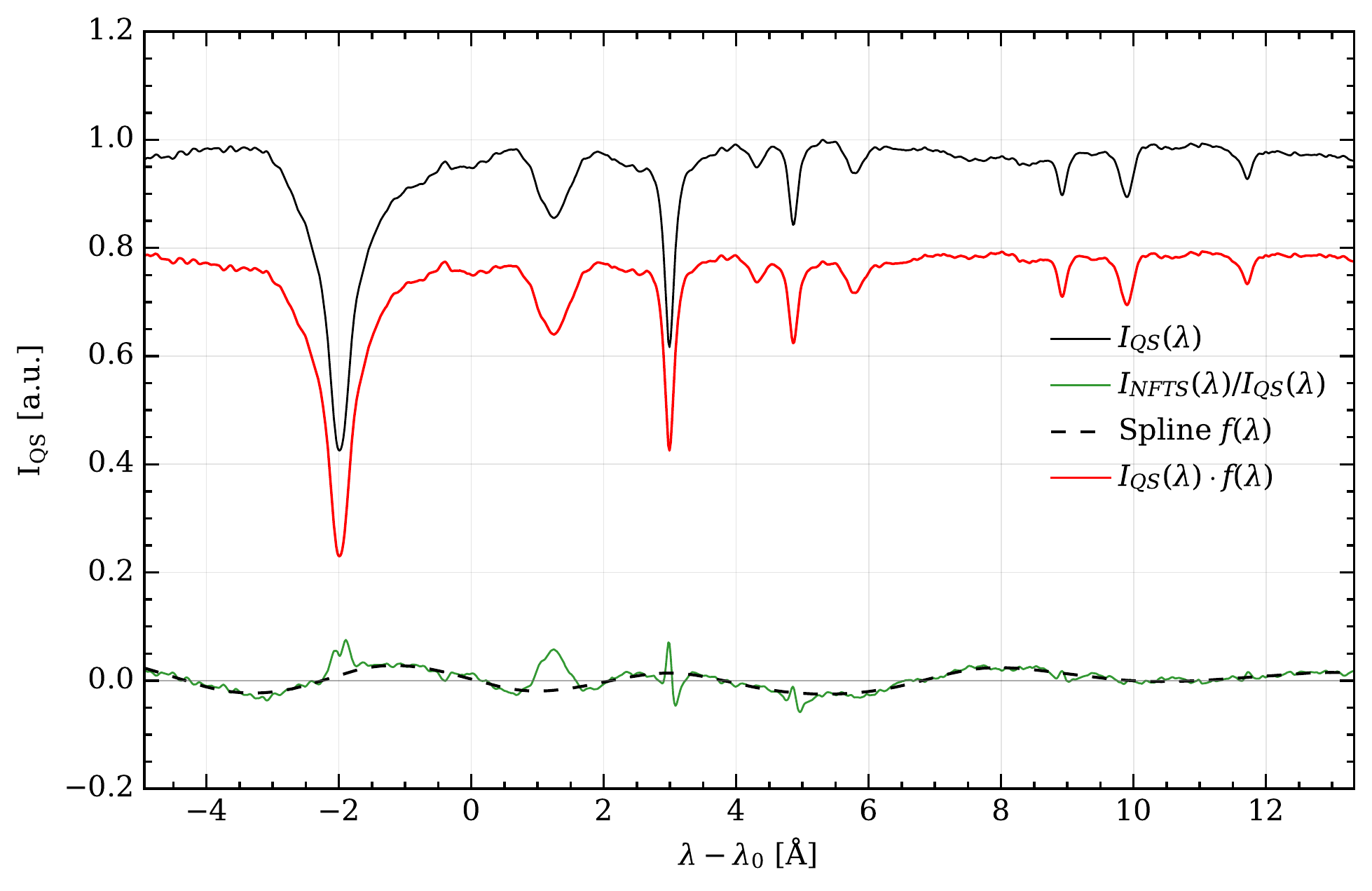}
\caption{Long-period fringes correction. The different parts of the correction process are shifted in the vertical axis to improve their visualization.}
\label{fig:fringes1}
\end{figure}

We create our $I_{QS}$ spectrum as an average of the quiet areas available in our observations to mimic the atlas. To compare both spectra, we performed a wavelength calibration using the photospheric lines of the atlas. Consequently, our wavelength reference (and LOS velocity) is the photospheric average motion. After the calibration, the dispersion obtained is 18.10$\pm$0.01\,m\AA\, pixel$^{-1}$.

The values of $\alpha$ and $\sigma$ are unknown and can be calculated with a least-square fit of the merit function:
\begin{equation}
\chi^2=\sum_i[I_{QS}(\lambda_i)-I_{\mathrm{NFTS}}(\lambda_i)]^2\cdot w_i \ ,
\label{eq:chiatlas}
\end{equation}
where $w_i$ is a parameter to reduce the weight of the wavelengths of the telluric lines. This weight is introduced because the telluric lines vary too much between observations. This weight is also applied to other spectral lines that could be highly dynamic such us the chromospheric \ion{He}{i}~10830\,\AA. Figure \ref{fig:fringes1} shows an example of the result when the values of $\alpha$ and $\sigma$ are calculated, and the fringe pattern is obtained from the ratio $I_{\mathrm{NFTS}}(\lambda)/I_{QS}(\lambda)$. This ratio $f(\lambda)$ is then smoothed or fitted with a spline in our case and used to correct all the pixels of the map by multiplying each of the Stokes parameters by $ f(\lambda)$. For the data analyzed in study, the values are $\alpha\sim15\%$ and $\sigma\simeq70$~m\AA.

For short-period fringes another method proved to be more efficient. These fringes are produced by interferences due to multiple reflections on flat surfaces in the optical path. They are often corrected using Fourier filtering, but this standard procedure does not achieve our requirements. We checked that this component is almost unpolarized and it is mainly seen in Stokes $I$. A first try to remove this component using Fourier or wavelet decomposition was unsuccessful. The reason is that the period of the fringes change along the spectrum and it is difficult to isolate them by simply using a threshold in the frequency domain. Thus, to remove this component we have used a technique based on the Relevant Vector Machine method\footnote{A python implementation of this tool can be found in \url{https://github.com/aasensio/rvm}.} \cite[RVM;][]{rvm2000} to decompose the original signal into two components: the clean spectra $I_0(\lambda)$ and the fringes $I_f(\lambda)$:
\begin{equation}
I(\lambda) = I_0(\lambda) + I_f(\lambda)\ .
\label{eq:franjas}
\end{equation}

\begin{figure}[!ht]
\centering
\includegraphics[width=\linewidth]{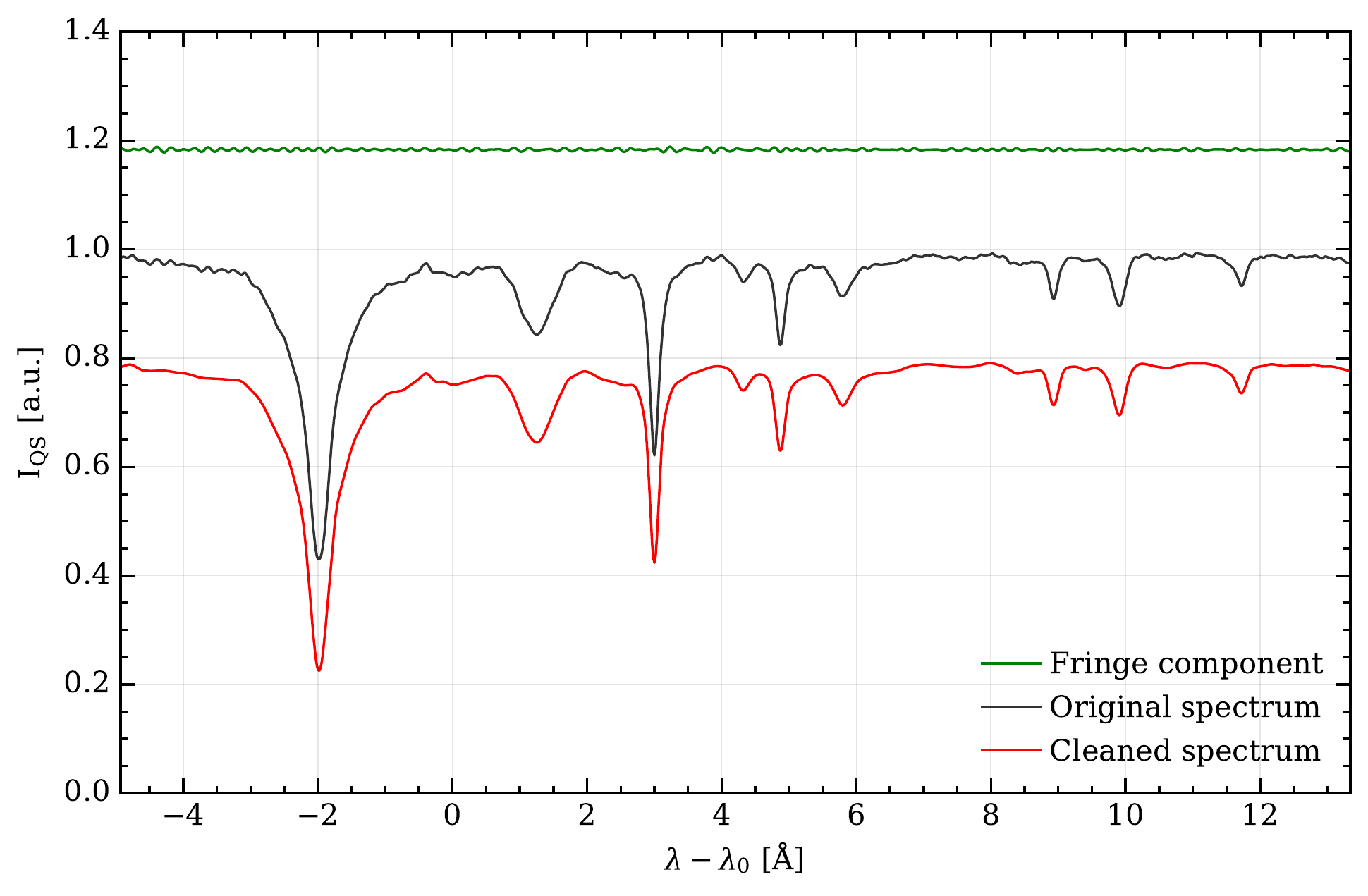}
\caption{Decomposition of the original signal using the RVM method into two components: the cleaned spectrum and the fringes. Each component was arbitrary shifted for a better visualization.}
\label{fig:rvm}
\end{figure}

Fringes have a narrow range of periods, not a very high amplitude and, most importantly, they are present over the entire spectral range, while spectral lines are present only in some parts of the spectrum. With these properties it seems convenient to use two dictionaries (in general, an overcomplete non-orthogonal basis set) with different properties  to decompose the two components. The first dictionary is made of Gaussian functions centered at several wavelengths $\lambda_j$ around the spectral lines with different widths $\Delta \lambda_j$. This dictionary is especially good at fitting the clean spectrum $I_0$ with just a few active elements. The second dictionary is made of sine and cosine functions of different periods $\omega_j$, similar to those of the fringes $I_f$. This dictionary is especially suited to fit the fringes with only a few active elements. Then, this problem can be mathematically described as a linear regression problem with $N$ coefficients $\vec{w} = (w_0, w_1, ..., w_G,..., w_N)$ such that:
\begin{align}
I(\lambda,\vec{w})&= w_0 + \sum^G_{j=1}w_j \exp\left[-\left(\frac{\lambda - \lambda_j}{\Delta \lambda_j}\right)^2\right]+ \nonumber \\ &+ \sum^{(N-G)/2}_{j=G+1} 
w_j \sin (\omega_j \lambda) + \sum^{N}_{j=(N-G)/2+1} w_j \cos (\omega_j \lambda)
\end{align}
where $G$ is the number of Gaussian functions. Note that half of the $N-G$ coefficients are associated with the sine function and the other half with the cosines. We have generated a dictionary with 3000 Gaussian functions and 400 sinusoidal functions with widths $\Delta\lambda$ between 5 and 10 pixels and frequencies $\omega$ between 0.4 and 0.8 pixel$^{-1}$. Given that the two dictionaries are very incoherent, it becomes hard to fit the spectral lines with the second dictionary and viceversa. {For the continuum level, the free parameter $w_0$ is used.} Therefore, the RVM method works by imposing a sparsity constraint on the coefficients of the dictionaries (forcing many elements of the dictionary to be inactive). With this, we managed to fit the spectrum at each position to separate the two contributions. Finally, the clean spectrum is obtained as the spectral line contribution $I_0(\lambda)$ without the fringes (see Fig. \ref{fig:rvm}). In this figure, this routine is able to separate fringes with 60 sinus and cosines and 100 Gaussian functions. Moreover, we do not have to worry about the Gaussians far from spectral lines because with the sparsity condition they are only activated if necessary. After this final step the data is ready to be analyzed.

No residual polarization is visible in the telluric lines. Thus, errors in the polarization calibration (that could appear as crosstalk) are significantly smaller than the Stokes $Q$ and $U$ signals that are the main focus of this work. 

To precisely determine the reference direction for $Q>0$ (an important parameter of our 
observation and inversion code), we tried to infer it directly from the observed maps. We 
found some discrepancies with the direction returned by the reduction pipeline. For the sake
of consistency we calculated this direction using the observations. Given that we observed a relatively round sunspot, we make the assumption that the magnetic field is oriented radially (in the sky plane) from the center of the umbra, something that is indeed observed as we show in the next sections. We find the reference direction for $Q>0$ by locating the direction where the integrated Stokes $Q$ profile is maximum, which should coincide with an almost zero Stokes $U$. This direction is different according to the rotation of each scan\footnote{In order to minimize the uncertainty in this angle, we apply the same procedure for each scan of the same day (not only these scans) and using both the silicon and helium line. The average value is used as an estimation of the reference direction. The average value is consistent among all scans with a dispersion of only three degrees.}. 
For the fifth scan, the $Q>0$ direction is at $23.7^\circ$ and the disk center is at a direction of $120.3^\circ$, both measured counter-clockwise from the horizontal (slit) axis (see Fig.~\ref{fig:MAPA2}). The spine of the filament is at $\sim42^\circ$ from the horizontal axis.

\begin{figure}[!]
\centering
\includegraphics[height=0.92\textheight]{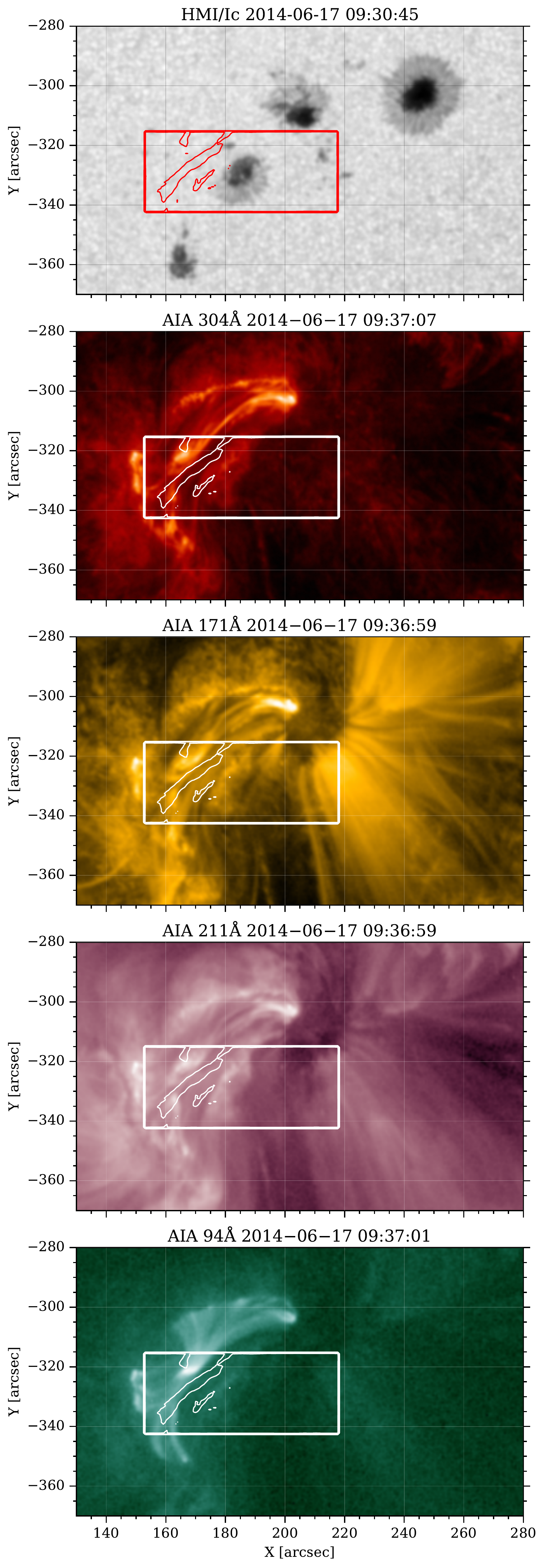}
\caption{The active region, NOAA 12087, with a contour showing our observed region above some AIA maps in different wavelengths indicated at the top of each panel. Despite the fact that the filament seems to be aligned with the coronal loops that are linking the two sunspots, we do not detected the filament in these maps.}
\label{fig:sdomaps}
\end{figure}

\begin{figure}[!ht]
\centering
\includegraphics[width=\linewidth]{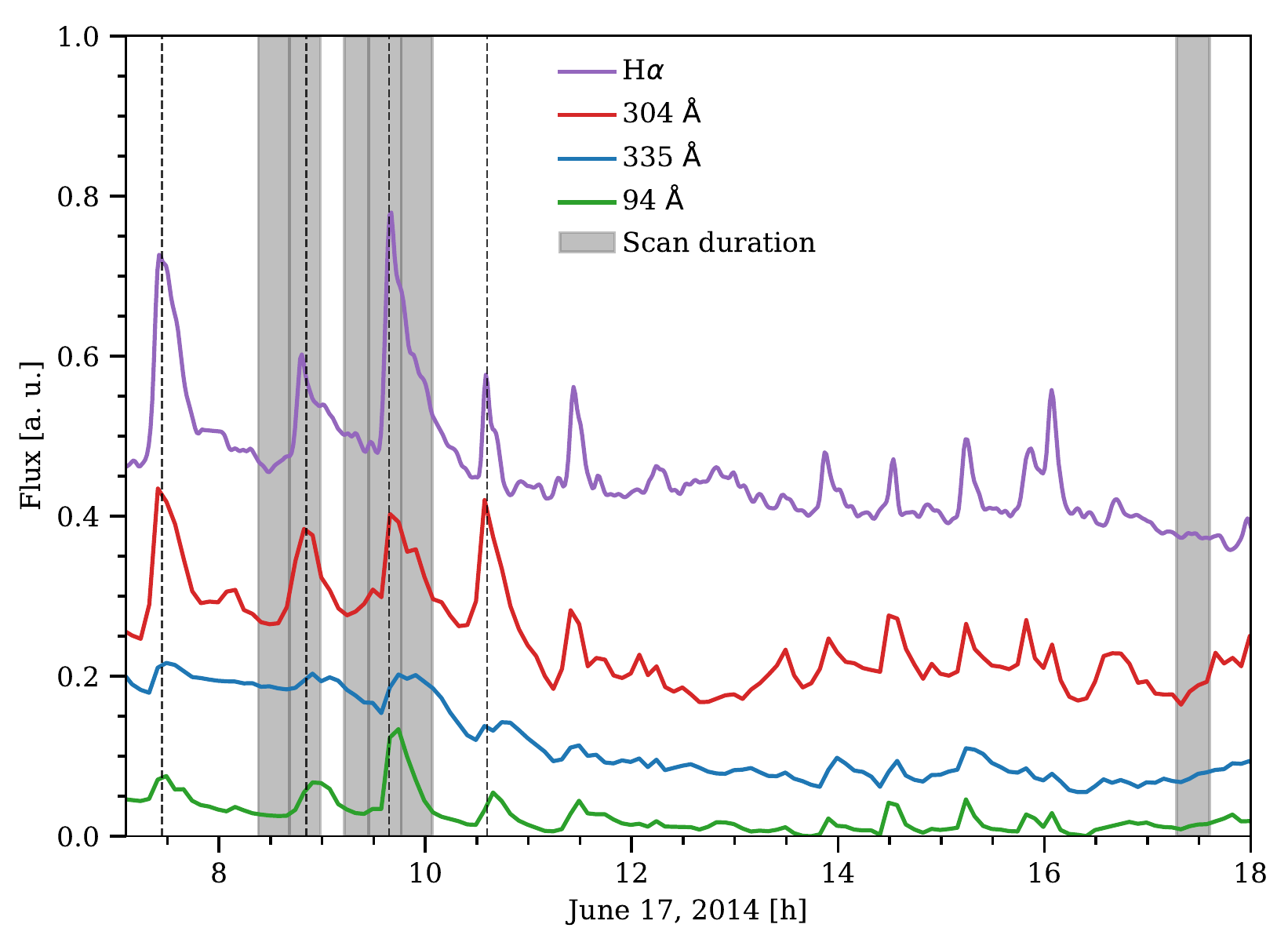}
\caption{Time evolution of the flux (in arbitrary units) integrated over the active region area for some filters. The six gray areas define the intervals of each scan ($\sim$15 minutes).}
\label{fig:flare131}
\end{figure}

\begin{figure*}[!ht]
\centering
\includegraphics[width=\linewidth]{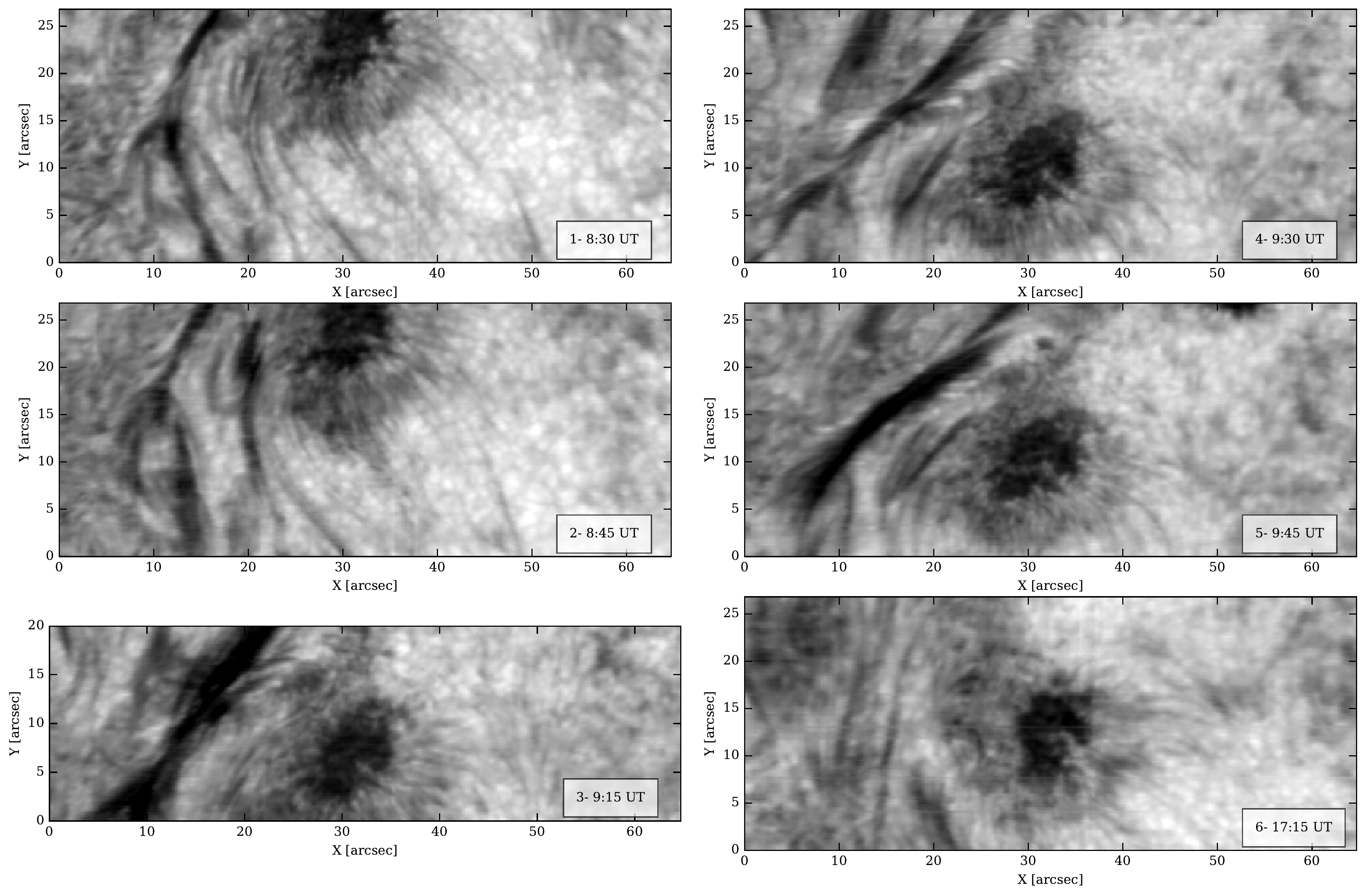}
\caption{This figure displays six reconstructed maps of the filament between 8:00 and 17:00 UT. The last scan was taken seven hours later showing how the filament has disappeared. These images are built with the intensity of the red component of the \ion{He}{i} multiplet at 10830.3\,\AA.}
\label{fig:evolution}
\end{figure*}

\subsection{Solar context}

In order to put the GRIS observations in context, we have made use of the data provided by some band-pass filters of the Atmospheric Imaging Assembly instrument \cite[AIA;][]{aia2012} and the Helioseismic and Magnetic Imager \cite[HMI;][]{hmi2012} on board NASA's Solar Dynamics Observatory \cite[SDO;][]{sdo2012}. The top panel of Fig. \ref{fig:sdomaps} shows an SDO/HMI image at the continuum of 6173~\AA\ of the NOAA 12087 region which we analyze here. The red box in Fig. \ref{fig:sdomaps} outlines the GRIS field of view (FOV). The HMI spatial resolution is about $\sim$1.2\arcsec\ \citep{yeo2014}. Due to the thermal structure of prominences, they can be seen in UV-EUV images. Several studies \citep{mackay2010,Parenti2012,McCauley2015} have demonstrated the usefulness of UV observations for plasma diagnostic in prominences. For that, the following panels display the SDO/AIA maps, from top to bottom: \ion{He}{ii}~304\,\AA, \ion{Fe}{ix}~171\,\AA, \ion{Fe}{xiv}~211\,\AA, and \ion{Fe}{xviii}~94\,\AA. The AIA spatial resolution is between 1.4$-$1.8\arcsec\ depending on the channel \citep{cali2012}.

Figure~\ref{fig:sdomaps} shows a very bright coronal loop linking the first and the third sunspot of the region (counting from left to right) in the AIA images. This brightness is enhanced in the feet and the apex of the loop. Our observed filament as seen in the core of \ion{He}{i} is overplotted with a white contour. The \ion{He}{i} absorption does not clearly correlate with any of the AIA channels. The reason might be that the filament we observed is low-lying, while the emission of higher coronal loops dominates the intensity measured in the AIA channels.

\subsection{Evolution of the filament}

During the same day of the observation several flares occurred. In particular, at the same time of our spectropolarimetric measurements, a GOES class B5\footnote{We remind the reader that the class of each flare is chosen according to their X-ray flux detected by the NOAA's GOES satellites in the wavelength range 1 to 8 \AA.} flare erupted in the central part of the active region inside our field of view. Since GOES observes the whole solar disk, we have to verify this, calculating the time evolution of the flux of the active region, which is shown in Fig.~\ref{fig:flare131}. We have used different filters such as the AIA channels at 304~\AA, 335~\AA, 94~\AA, and H$\alpha$ obtained by the GONG network \citep{hill1994}.

Figure \ref{fig:flare131} displays a clear peak during the fourth scan. In fact, the energy is released close to the central part of the filament but not exactly above the filament. Figure \ref{fig:sdomaps} shows this compact bright point at (170\arcsec,\,-320\arcsec). We see a sudden and intense variation in brightness from 09:35 to 9:46UT in the 94~\AA\ and 304~\AA\ channels and a weaker brightening in the remaining filters. We do not detect either emission intensity profiles of the \ion{He}{i} or a significant decrease of the intensity profiles during the flare, in contrast to what have been found other studies \citep{sasso2011,kuckein2015,judge2015}, possibly because our flare is one or two orders of magnitude less energetic than theirs in X-rays. Another possibility is that the flare occurs at higher layers above the filament, and the energy is redirected towards both feet of the big loop, as some brightening at (200\arcsec,\,-300\arcsec) and (150\arcsec,\,-320\arcsec) seem to indicate. This option is also compatible with the fact that while the absorption of the filament changes dramatically during its lifetime, the polarization characteristics (amplitude and distribution of the signals following the filament) remain almost constant. This may indicate that the magnetic topology does not change much while changes in the temperature and density are causing changes in the line opacity.

We have also analyzed how the filament shape changes during its lifetime. Figure \ref{fig:evolution} shows monochromatic images of the core of the \ion{He}{i} line reconstructed from all the observed scans. In the first two scans, the filament is made of several thin threads. At 9:15 UT during the third scan, the filament has a compact body with the largest width of the time series. Later, it changes its shape again and we argue that the two flares at 8:45 UT and 9:30 UT are correlated with the time steps in which the filament is more diffuse. The energy released could evaporate or destabilize the filament structure seen in the \ion{He}{i} triplet. After seven B-class flares, at 17:15 UT the filament has almost vanished. Although we do not see a very clear relation between the filament and the flares, if reconnection events are taking place, a reconfiguration of the magnetic field of the environment could be the reason why the evolution of the filament is so fast.

\begin{figure*}[!ht]
\centering
\includegraphics[width=\linewidth]{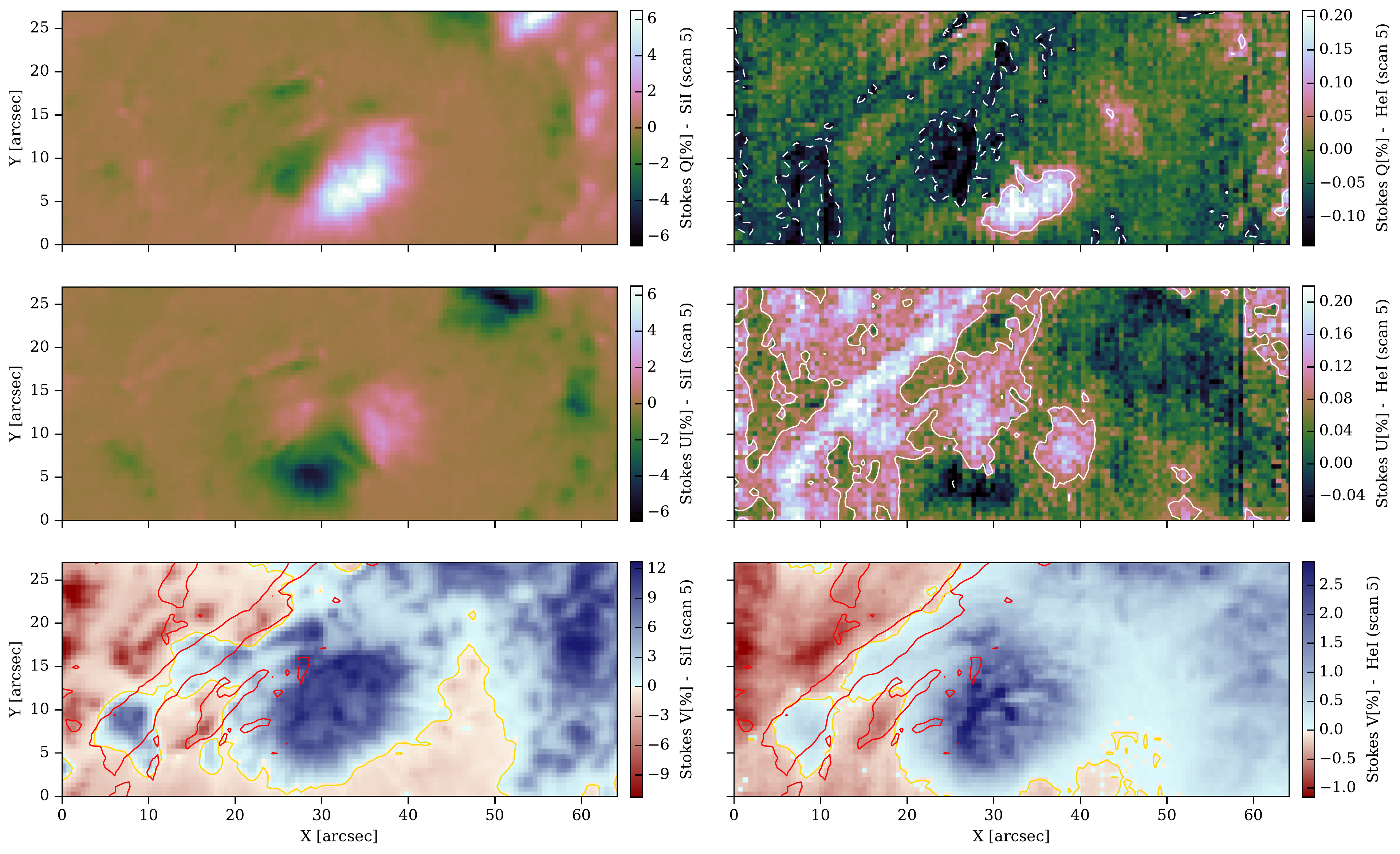}
\caption{Stokes $Q(\lambda_{0})/I_c$, $U(\lambda_{0})/I_c$ and the amplitude of the red lobe ($\lambda_{R}$) of Stokes $V(\lambda_{R})/I_c$ for the \ion{Si}{i} line at $10827.1$\AA\ (left column) and the red component of the \ion{He}{i} triplet at $10830.3$\,\AA\ (right column). The white solid (dashed) contour shows the positive (negative) signals above three times the noise level. Red contours at 0.7$I_c$ show where the filament is located.}
\label{fig:PolMap5}
\end{figure*}

\begin{figure*}[!ht]
\centering
\includegraphics[width=\linewidth]{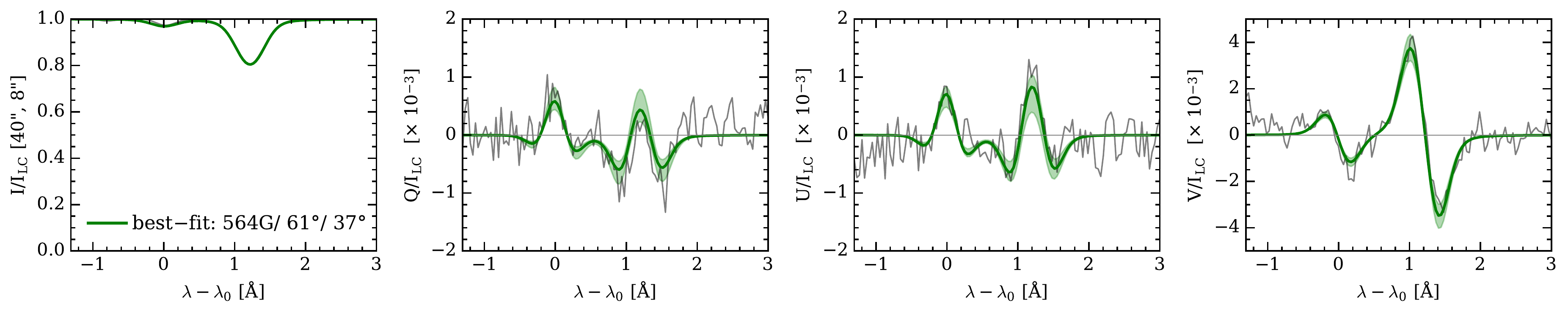}
\includegraphics[width=\linewidth]{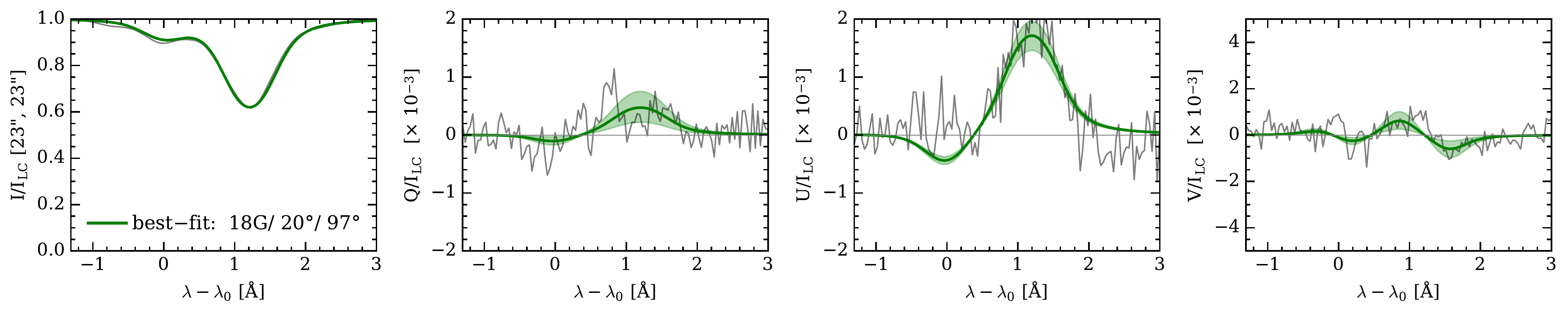}
\caption{{Observed (grey) and best-fitted (green) \ion{He}{i} 10830\,\AA\ triplet profiles corresponding to a typical penumbral (upper row) and a filament profile (lower row). The coordinates of each pixel in the map are indicated in the y-label. A shadowed green area is displayed indicating the range of solutions compatible with the profiles obtained with the method described in Sec.~\ref{sec:bayesian}}. In the left column panels the parameters of the best fit are displayed. The symbol $I_{LC}$ is the local continuum intensity.}
\label{fig:zoology}
\end{figure*}

\begin{figure}[!ht]
\centering
\includegraphics[width=\linewidth]{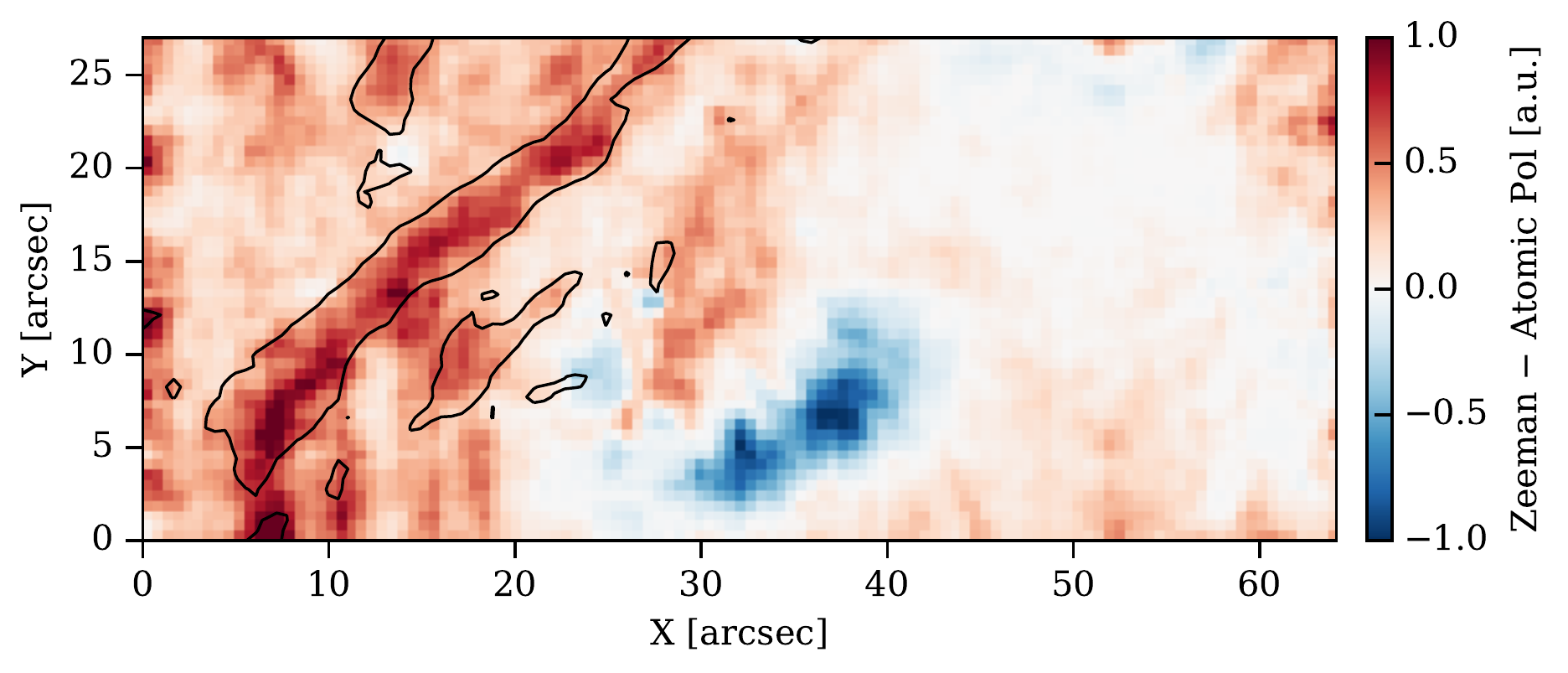}
\caption{A classification of the Stokes $Q$ and $U$ profiles of the region. Zeeman-like profiles are drawn in blue and Hanle-like profiles in red. Polarization signals are weighted by its strength, with the white color indicating low polarization signals where the shape is unclear. A contour at 0.7$I_c$ is drawn to show the filament location.}
\label{fig:classification}
\end{figure}

\section{Analysis of the polarization signals}

\subsection{Polarization maps}
We start by analyzing the spatial distribution of the polarization signals. Since all the scans have similar polarization properties and our aim is the study of the filament itself, we have focused our study in the fifth scan, where the longest and most compact structure is seen. Figure~\ref{fig:PolMap5} displays the Stokes $Q(\lambda_{0})/I_c$ and $U(\lambda_{0})/I_c$ at the center wavelength of the line ($\lambda_{0}$) and the amplitude of the red lobe at $\lambda_{R}$ of Stokes $V(\lambda_{R})/I_c$ for the \ion{Si}{i} line at $10827.1$\,\AA\ (in the left column) and for the red component of the \ion{He}{i} triplet at $10830.3$\,\AA\ (in the right column). White contours show the signal above three times the noise level. Solid line contours indicate positive signals while negative signals are displayed with dashed line contours. Red contours display the location of the absorption features (i.e., the filament).
The linear polarization maps of the silicon line show strong signals in the penumbra of the sunspot. Although the Stokes $Q$ amplitude of the \ion{He}{i} triplet does not exhibit an evident correlation with the filament, we can detect a strong signal coincident with the body of the filament in Stokes $U$. The average Stokes $U$ signal at the line core within the filament is around $1.5\times10^{-3}I_c$. 

{Finally, in the last row of Fig.~\ref{fig:PolMap5} we display the Stokes $V$ maps for the \ion{Si}{i} and \ion{He}{i} lines. The \ion{Si}{i} Stokes $V$ map shows the polarity of the magnetic field in the whole area (pointing towards the observer in blue and away in red).  Interestingly, the \ion{He}{i} Stokes $V$ map does not show any correlation with the filament at all and it is very similar to the photospheric map but more diffuse. This fact seems to suggest that the measured circular polarization is dominated by the polarization generated in the magnetized region below and not from the filament. This is not surprising since the fact that the filament could be transparent to polarized radiation originating in the underlying magnetized chromosphere has already been shown by \cite{diaz2016}.}

\subsection{General description of the helium profiles}

In this section we describe the morphology of the observed \ion{He}{i} Stokes profiles. In general, it is possible to have a quite precise idea of which is the main mechanism dominating the polarization signals in each region by the shape of the Stokes parameters. Anisotropic radiative pumping, the Hanle effect, and the Zeeman effect work together to shape the polarization of the \ion{He}{i} triplet. Unless specified, when we refer to each Stokes parameter, they will be normalized to their local continuum, $I_{LC}$, computed in the same pixel and in the nearby spectral continuum region. More details will be given in Sec.~\ref{sec:hazel}.

The strong absorption observed in Fig.~\ref{fig:MAPA2} cause broader and deeper Stokes $I$ profiles as compared to the quiet regions. For a pixel inside the filament we observe signatures of scattering polarization (and likely Hanle effect) in Stokes $Q$ and $U$. As an example, the lower four panels of Fig.~\ref{fig:zoology} display typical Stokes profiles of that area. The \ion{He}{i} profiles exhibit the usual single-lobed symmetric shape in linear polarization \citep{TrujilloBueno2002}. Contrary to the Zeeman effect, the presence of a magnetic field is not necessary in general to produce polarization signals, which are generated by scattering processes. However, the presence of a magnetic field affecting the polarization through the Hanle effect can be inferred from Fig.~\ref{fig:PolMap5}. In absence of magnetic fields, {due to geometrical considerations we expect all the linear polarization to be parallel to the closest limb}\footnote{This result is strictly true only if the radiative transfer inside the filament can be neglected and the radiation field can be considered axial symmetric.} \citep{bommier1989,TrujilloBueno2002,milic2016}. Since our Stokes $Q>0$ reference direction is defined almost along this direction (parallel to the limb), we would expect the linear polarization of the filament to be observed only in Stokes $Q$. However, Fig \ref{fig:PolMap5} shows that the linear polarization signal is fundamentally contained in Stokes $U$, while Stokes $Q$ is barely detectable. This modification of the polarization is a clear signature of the magnetic field acting via the Hanle effect \citep{landi_landolfi04}.

We also note from the lower row of Fig.~\ref{fig:zoology} that both Stokes $Q$ and $U$ profiles extracted from the filament show polarization in the blue component ($\lambda_0$ in this figure) of the \ion{He}{i} multiplet. Since the upper level of the blue level $(^3P_0)$ has angular momentum $J=0$, it cannot be polarized and the emitted radiation has to be unpolarized. The polarization is then a consequence of dichroism because the lower level ($^3S_1$) has angular momentum $J=1$ and can be polarized \citep{TrujilloBueno2002}. {We expect the signals in the blue and red components to have opposite signs given their polarizability, which is defined as a factor that depends on the quantum numbers of the transition and accounts for the transformation of the properties of the radiation field into atomic populations}. This is not the general behaviour in our observations. We discuss this issue in detail in the second part of the series. For the moment, and given the small amplitude of the blue component, we neglect its effect on the single component inversion presented in this first analysis.

Far from the filament and close to the sunspot the linear polarization signals exhibit the typical Zeeman profiles in Stokes $Q$ and $U$. These profiles are characterized by three lobes in the linear polarization {as we see from the profiles shown in the upper row of Fig.~\ref{fig:zoology} that have been extracted from the penumbra of the sunspot}. This region is Zeeman-dominated and, consequently, for this spectral line, the magnetic field strengths are expected to be on the hG--kG regime.

In order to visually localize each profile in the observed FOV and, consequently, which mechanism dominates the polarization signals, we have classified the Stokes $Q$ and $U$ profiles using a \textit{k-means} clustering algorithm \citep{kmean79}. Figure \ref{fig:classification} shows the location of single-lobed profiles in red and three-lobed profiles in blue. The specific value of each pixel is set by the class-value $C$ ($+1$ for single-lobed and $-1$ for three-lobed profiles) multiplied by the absolute value of its linear polarization signal: $C_Q\cdot|Q(\lambda_0)| + C_U\cdot|U(\lambda_0)|$, where $C_Q=\pm 1$ and $C_U=\pm 1$ depending on the shape of the profile in this Stokes parameter, and $\lambda_0$ is the center of the red component of the \ion{He}{i} triplet at $10830.3$\,\AA. This product enhances those pixels where the classification is reliable and sets closer to zero those areas where the profile is very noisy. Finally, the result is normalized to the maximum value of the map.

Concerning Stokes $V$, the whole map shows the typical antisymmetric circular polarization profile generated by the longitudinal Zeeman effect. We note that the Stokes $V$ signal in the filament is weaker than in the sunspot. Sometimes, the regular shape is hard to distinguish because of the noise, but we can safely say that no orientation-induced signals like those of \cite{MartinezGonzalez2012} are found.

\subsection{Inference of dynamic and magnetic properties}

\subsubsection{Photospheric inference}\label{sec:sir}

The \ion{Si}{i} 10827\,\AA\ line in the observed spectral range allows us to extract the thermodynamic and magnetic information of the photosphere below the filament. As described before, other photospheric \ion{Ca}{i} lines are located a few Angstroms  towards the red \citep{joshi2016,tobias2016}, but the \ion{Si}{i} 10827\,\AA\ is enough to have an estimation of the photospheric magnetic field. Some studies have demonstrated that the formation of the \ion{Si}{i} line presents some departures from local thermodynamic equilibrium \citep{bard2008,Shchukina2017}, with the core of the silicon line originating in the upper photosphere. Non-LTE effects are certainly non-negligible in the estimation of the temperature stratification, but the impact on the inferred magnetic field is  limited \citep{kuckein2012}. For this reason, we use an LTE inversion code which is less accurate but faster than non-LTE inversion codes.

The \ion{Si}{i} line has a moderate magnetic sensitivity with an effective Land\'e factor of $\bar{g}=1.5$. We have checked with the Stokes $V$ response functions (RFs) to the magnetic field that the highest sensitivity is found close to $\log(\tau)=-2.2$ \citep{kuckein2012,tobias2016}, where $\tau$ corresponds to the  continuum optical depth at 5000\,\AA. All the physical magnitudes inferred from the \ion{Si}{i} line are displayed in the following at that height. The \ion{Si}{i} 10827\,\AA\ line was inverted using the \emph{Stokes Inversion based on Response functions} code \citep[\Sir,][]{sir1992}, that assumes LTE and hydrostatic equilibrium to solve the radiative transfer equation. We carried out a pixel-by-pixel inversion with an initialization of the inclination of the magnetic field given by the Stokes $V$ polarity (45$^\circ$ if the blue lobe is positive and 135$^\circ$ if it is negative).

\Sir\ inversions require an initial guess for the atmospheric model. We choose the \emph{Harvard-Smithsonian Reference model Atmosphere} \citep[HSRA,][]{HSRA1971}, covering the optical depth range $-5.0 < \log(\tau) < 1.4$. We add to the HSRA model a magnetic field strength of 500\,G. The inversions were performed in three cycles. The temperature stratification $T(\tau)$ was inverted using 2,\,3, and 5 nodes in each cycle, and two nodes were used for magnetic field strength $B(\tau)$, magnetic field inclination $\Theta_B(\tau)$, and line-of-sight velocity v$_{LOS}(\tau)$. The magnetic field azimuth $\Phi_B$ was set constant with height. The solar abundances were taken from \cite{Asplund2009}, while the atomic data for the \ion{Si}{i} 10827\,\AA\ line was obtained from \cite{Borrero2003}. The results of the inversion will be shown in Sec.~\ref{sec:inversion}.

\subsubsection{Chromospheric inference}\label{sec:hazel}

For extracting physical information from the chromospheric \ion{He}{i} 10830\,\AA\ triplet we have used the inversion code \emph{Hanle and Zeeman Light} \cite[\Hazel,][]{asensio2008}. This code is able to infer the magnetic field vector from the emergent Stokes profiles considering the scattering  polarization and the joint action of the Hanle and Zeeman effects. It assumes a slab of constant physical properties located at a certain height over the visible solar surface, illuminated by the photospheric continuum and permeated by a deterministic magnetic field. This slab model is fully described using 9 parameters: the optical depth $\tau_R$ in the core of the red component, the damping parameter, the Doppler width, the line-of-sight velocity of the plasma, the height, the heliocentric angle, and the three components of the magnetic field vector. The radiation anisotropy due to the center-to-limb variation is taken into account. \Hazel\ modifies iteratively the parameters of the slab model in order to produce a set of synthetic Stokes spectra that best match the observed ones using a least-squares algorithm. 

As suggested in \cite{asensio2008}, we used a two-step procedure in the inversion process. First we obtain the optical depth, the velocity in the LOS, and the Doppler width from Stokes $I$. Then, we infer the magnetic field vector using all Stokes parameters, keeping the already inverted parameters fixed. To fully characterize the incident radiation field, we need to determine the height $h$ of the filament above the solar surface. In our case, this height cannot be measured since we see the filament from above, as the LOS is almost perpendicular to the solar surface at that point. For this reason, we have used a typical value of $h=3\arcsec\simeq 2200$~km above the surface \citep{kuckein2012,YellesChaouche2012,mackay2010} for the whole FOV, as we do not know a priori if the filament is located much higher above the solar surface than the rest of the FOV. To validate our choice, we have used the projected distance between the polarity inversion line obtained from the \ion{Si}{i} and \ion{He}{i} lines. Under the assumption that the PIL lies roughly at the same local vertical position, one can use the projected distance in the sky and associate it to a difference in height. In our observation we find a difference of $\sim$1.5\,\arcsec, which corresponds to $\sim 2000$\,km using Eq.\,1 of \cite{joshi2016}. Assuming an average formation height for the \ion{Si}{i} line of $\sim300$\,km \citep{Shchukina2017}, we find that the filament is lying at $h\sim 2300$\,km above the surface  roughly similar to the height that we use. We extend the discussion on the height of the filament in Sec.~\ref{sec:height}.

\begin{figure}[!ht]
\centering
\includegraphics[width=0.95\linewidth]{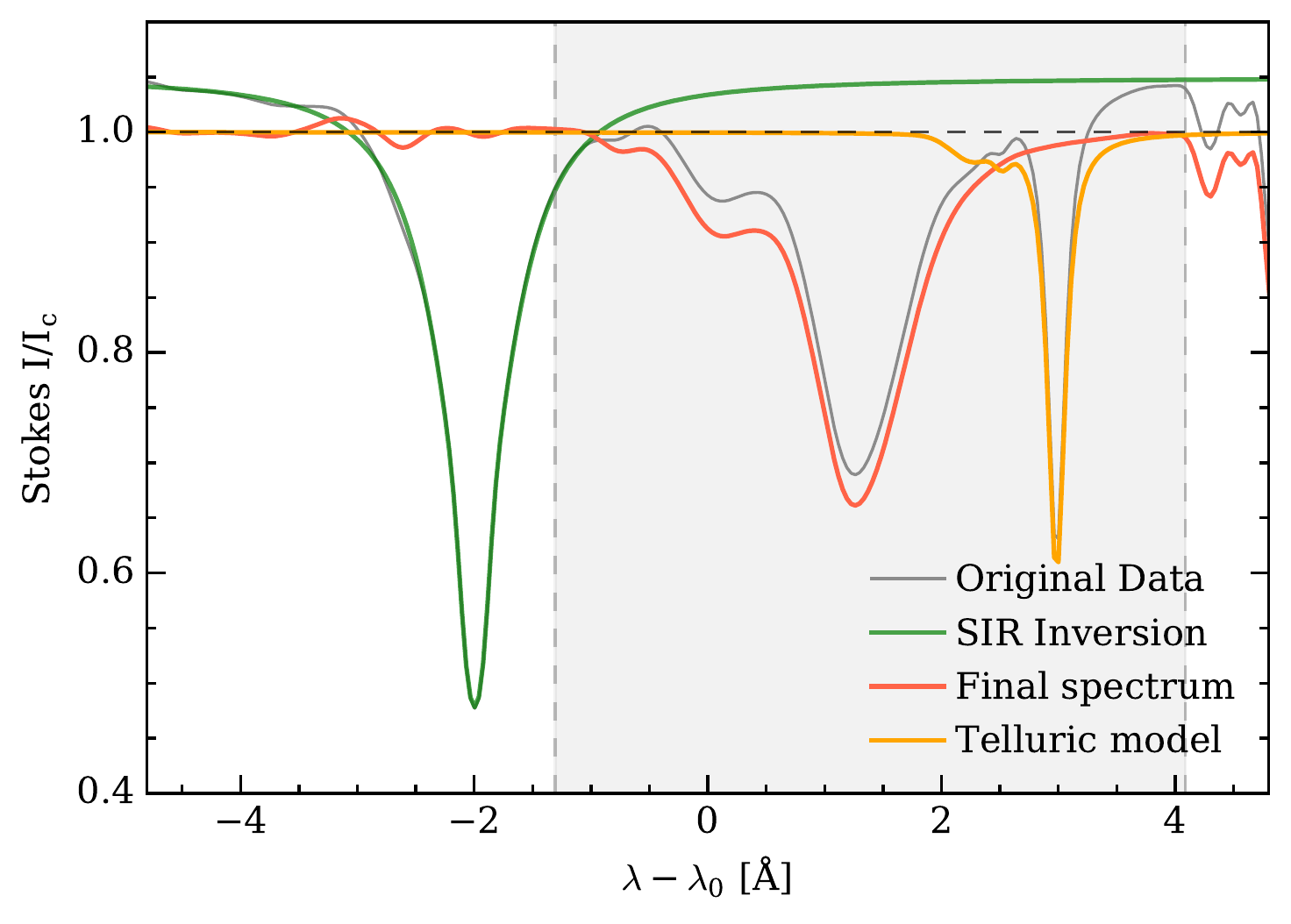}
\caption{Example of the normalization procedure of the helium triplet. The shadowed region is the spectral range inverted with \Hazel.}
\label{fig:normalizacion}
\end{figure}

While the anisotropy of the radiation field used in \Hazel\ to compute the density matrix is computed using the center-to-limb variation of the photospheric continuum \citep{cox00}, the boundary condition used for the formal solution of the radiative transfer equation to get the emergent Stokes profiles can change from pixel to pixel and it has to be provided to \Hazel. Moreover, the data must be normalized to its local continuum $I_{LC}$. This is sometimes  difficult because the nearby \ion{Si}{i} line has strong wings. Therefore, we have used the inversion of \Sir\ to model the \ion{Si}{i} line profile. We extend the synthetic wing towards the \ion{He}{i} line and then normalize the spectrum by the synthetic profile. The nearby telluric lines have been also removed by modeling them wherever they are not blended with the \ion{He}{i} triplet. We have modeled the variation with time of the telluric lines by fitting a Voigt profile. After this correction  we are ready to invert the data. Figure~\ref{fig:normalizacion} displays an example of this normalization procedure showing the important contributions of the silicon and telluric lines. After applying this normalization, the filament absorption remains visible in the core intensity map, while the granulation pattern and the sunspot disappear.

To fix the reference system for these observations, we have chosen that the projection of the $X$ axis in the plane of the sky is in the radial direction pointing towards the disk center (i.e., the local azimuth angle is $\phi=180^\circ$). The angle between the radial direction ($X$ axis) and the $Q>0$ reference of the polarimeter \citep[$e_{1}$, see ][]{landi_landolfi04,asensio2008} is $\gamma\sim83^\circ$ for the fifth scan\footnote{See the manual of \Hazel\ for a detailed explanation of how to carefully calculate these angles: \url{http://aasensio.github.io/hazel/refsys.html}}. 

A reliable inversion of the \ion{He}{i} multiplet required the aforementioned spatial and spectral binning. Even after the rebinning, the polarimetric signals are only few times above the noise. Having a reliable determination of Stokes $V$ turns out to be fundamental to correctly infer the strength of the magnetic field. The reason is that between 10 and 100\,G the \ion{He}{i} 10830\,\AA\ multiplet is in the Hanle saturation regime, where the linear polarization is only sensitive to the direction of the magnetic field and not to its strength. Therefore, the only way to infer the full magnetic field vector is through the full Stokes vector. The results of the inversion will be shown in the next section.

\begin{figure*}
\centering
\includegraphics[width=1.0\linewidth]{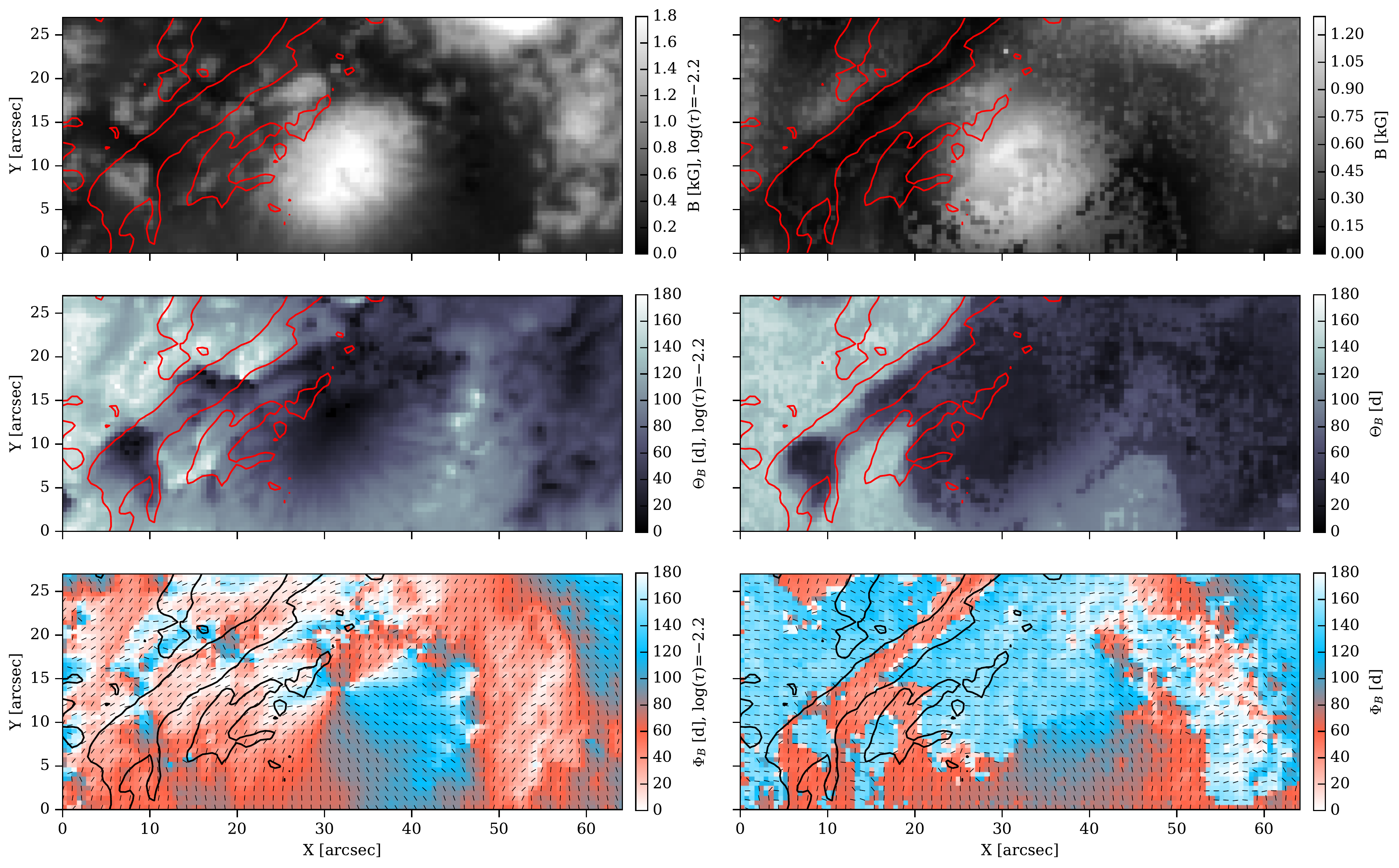}
\caption{Magnetic field strength (top row), inclination (middle row) and azimuth maps (bottom row) inferred from the inversion of both spectral lines in the line-of-sight reference frame. The left column shows the \Sir\ inversion of the \ion{Si}{i} line, while the right column shows the \Hazel\ inversion of the \ion{He}{i} multiplet. Contours include all the points with an intensity lower than 0.7$I_c$, delineating the location of the filament.}
\label{fig:inversionA}
\end{figure*}

\section{Inversion results}\label{sec:inversion}

\subsection{Magnetic field vector in the LOS reference frame}
\label{sec:Blos10830}

Figure~\ref{fig:inversionA} shows the magnetic field strength, inclination, azimuth, and LOS velocity maps for the fifth scan, all of them in the LOS frame $(\Theta_B,\Phi_B)$. We note that the azimuth and inclination of the \ion{He}{i} line have been transformed from the local reference frame, the one used by \Hazel, to the LOS reference frame to properly compare the silicon and the helium inversion. The photospheric magnetic field strength map shows a $\sim$2\,kG sunspot with only half of the penumbra visible. In the chromosphere, the magnetic field strength of the sunspot reaches values of $\sim$1\,kG. 

The magnetic field strength map shows that the lowest values are associated with the location of the filament, specially in the chromosphere. The magnetic field strength of the area coinciding with the filament shows values around 50--150\,G, similar to those quoted by \cite{sasso2014}. The LOS inclination map shows how the FOV is split into two polarities with a very narrow PIL following the filament in the chromosphere, but not so well aligned  in the photosphere. 
Concerning the azimuth, and eluding for the moment the fact that other ambiguous solutions exist, the results show a penumbra with a clearly radial magnetic field. We can affirm this with confidence because we are only affected by the 180$^\circ$ ambiguity in this Zeeman-dominated region. The azimuth in the umbra is not reliably retrieved for \ion{He}{i} since the magnetic field is more vertical and Stokes $Q$ and $U$ have lower amplitudes. Given the multiple potential solutions for the azimuth in the chromosphere, one should notice that it is possible to find fields compatible with the observations that are parallel or perpendicular to the filament. We study this problem  with more detail in Sec. \ref{sec:ambiguities} and \ref{sec:bayesian}.

\subsection{Azimuth in the local frame}

Because of the relative simplicity of solving the 180$^\circ$ ambiguity in the interior of the sunspot, we have decided to transform the photospheric azimuths given in the LOS by \Sir\ to the local reference system. To perform this disambiguation we used a simple potential field extrapolation using the vertical component of the magnetic field and selected the inferred solution at each pixel that is more aligned with the potential field \citep[also called Acute-Angle-Comparison,][]{Metcalf2006}. The local \ion{Si}{i} azimuth map is displayed in the upper part of the Fig. \ref{fig:local}. As expected, the magnetic field vector in the sunspot is pointing radially outwards from the center of the umbra. In the area where the filament is located the azimuth is apparently aligned with it. However, we note that the potential field extrapolation is probably not the best approximation to the field topology due to the complexity of this region, which might introduce errors in the disambiguation.

The local azimuth map inferred by \Hazel\ is displayed in the middle panel of Fig.~\ref{fig:local}. In this case, the azimuth map is not disambiguated. An interesting observation is that there is a clear correlation between the azimuth and the inclination map displayed in Fig. \ref{fig:inversionA}. While in the LOS reference frame the longitudinal magnetic field information is purely contained in Stokes $V$ and the transversal information lies in Stokes $Q$ and $U$, in the local reference frame (the observation is not at disk center)  Stokes $V$ also provides information about the magnetic field parallel to the solar surface.

A crucial conclusion from Fig.~\ref{fig:local} is that the chromospheric azimuth inside the filament is not smooth, i.e., we witness strong variations in the local azimuth (close to 180$^\circ$)  between nearby pixels at the PIL. While Stokes $Q$ and $U$ have very similar shapes in the whole filament, Stokes $V$ changes sign and it is responsible of these changes in the azimuth. Apart from other compatible solutions, this unphysical situation could arise if the measured Stokes $V$ does not come from the filament itself, but from the active region underneath, something that is explored in the second part of this study. In such situation, one would be using \Hazel\ to simultaneously interpret circular and linear polarization profiles as coming from the same atmospheric region in the single component model.

\subsection{Ambiguous solutions}\label{sec:ambiguities}
The main reason why we have not disambiguated the local azimuth map
in the chromosphere is that there are many possible ambiguous solutions. In fact, the number of these ambiguous solutions depend on the specific regime of the magnetic field and on the scattering geometry. These ambiguities produce equally good fits to the
observed Stokes profiles and they are then impossible to distinguish 
based on pure spectropolarimetric considerations.
\begin{figure}[!ht]
\centering
\includegraphics[width=1.0\linewidth]{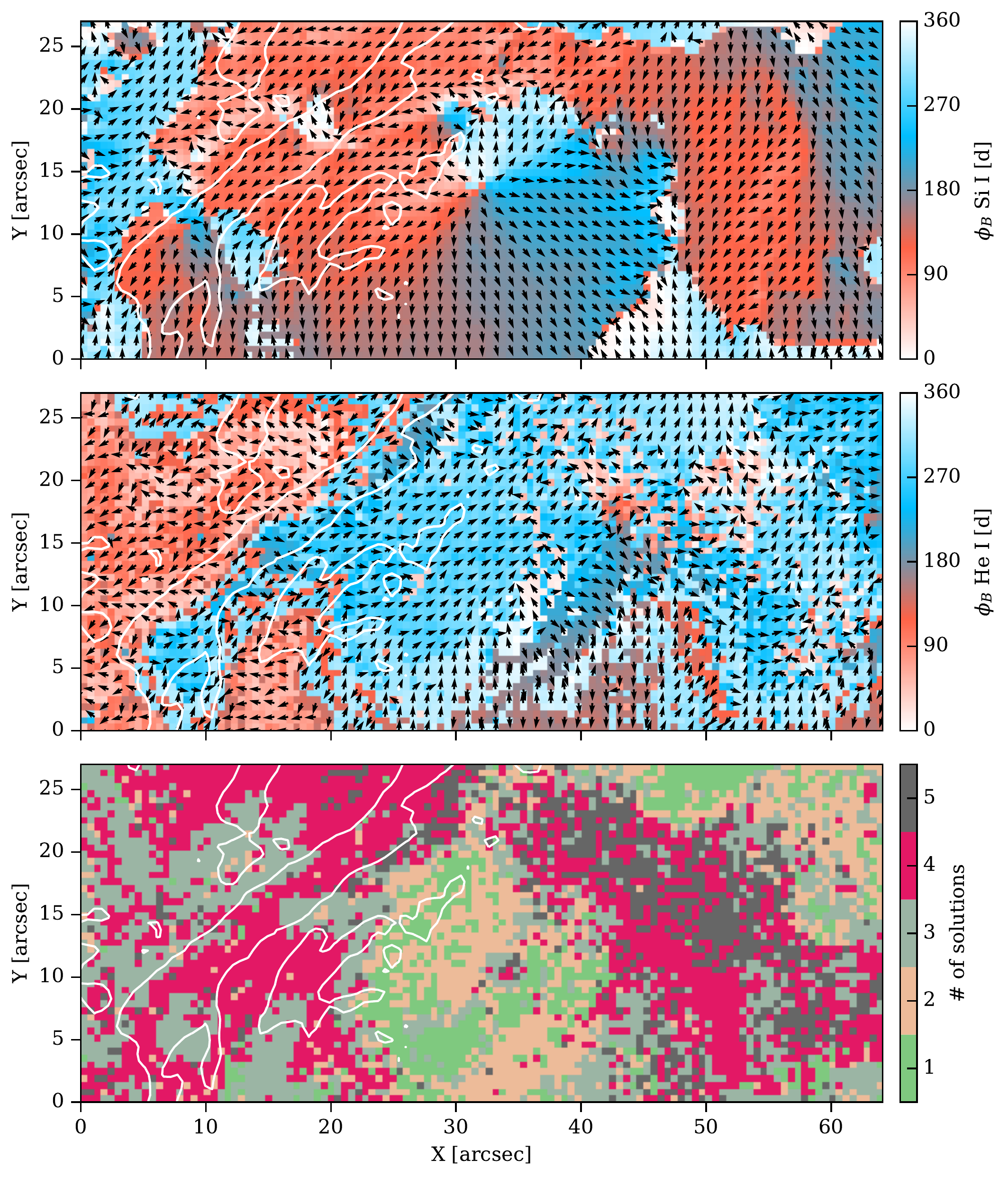}
\caption{Upper panel: local azimuth map for the \ion{Si}{i} inversion obtained after a transformation from the LOS frame using a potential field approximation. Middle panel: local azimuth map for the \ion{He}{i} inversion. Lower panel: number of compatible solutions for each pixel.}
\label{fig:local}
\end{figure}

In the plane of the sky the determination of the orientation of the magnetic field vector from the polarization profiles in the Hanle saturated regime suffers from two known ambiguities: the 180$^\circ$ ambiguity (fields with $\Phi_B$ and $\Phi_B+180^\circ$ azimuths give the same signal) and the Van Vleck or 90$^\circ$ ambiguity (fields with $\Phi_B$ and $\Phi_B\pm90^\circ$ give the same signals). One can estimate the other ambiguity solutions of the Hanle effect in the saturation regime with the recipe of \cite{marian2015}. They used a two-level atom with $J_u=0$ and $J_l=1$ in the optically thin limit. This is a relatively good approximation for the \ion{He}{i} 10830\,\AA\ blue component, but not for the red component. For that, with this first guess, one have to refine the solution with \Hazel. We apply this procedure for all pixels and display the number of compatible solutions per pixel in the lower panel of Fig.~\ref{fig:local}. In our field of view, similar to what was found by \cite{schad2015}, there is a non-negligible fraction of pixels with only one possible solution. The number of solutions go up to two inside both sunspots (one in the center and another in the upper right corner). They are due to the well-known 180$^\circ$ ambiguity. Away from the sunspot, where the magnetic field is weaker, more than 3 solutions appear. More than 5 solutions are found in regions  where the signal is very low, due to the noise (all of them labeled as 5). Finally, in the region of the filament we find between three and four compatible solutions.

The geometry of the observation turns out to be important because it gives us the advantage, in some pixels, of having only one compatible solution. The areas with only one possible solution correspond to regions in which the Hanle and Zeeman effects act together. We defer the detailed study of these signals to Sec.~\ref{sec:multimodal}.

\subsection{Line-of-sight velocities}\label{subsec:10830vlos}

\begin{figure}[!ht]
\centering
\includegraphics[width=1.0\linewidth]{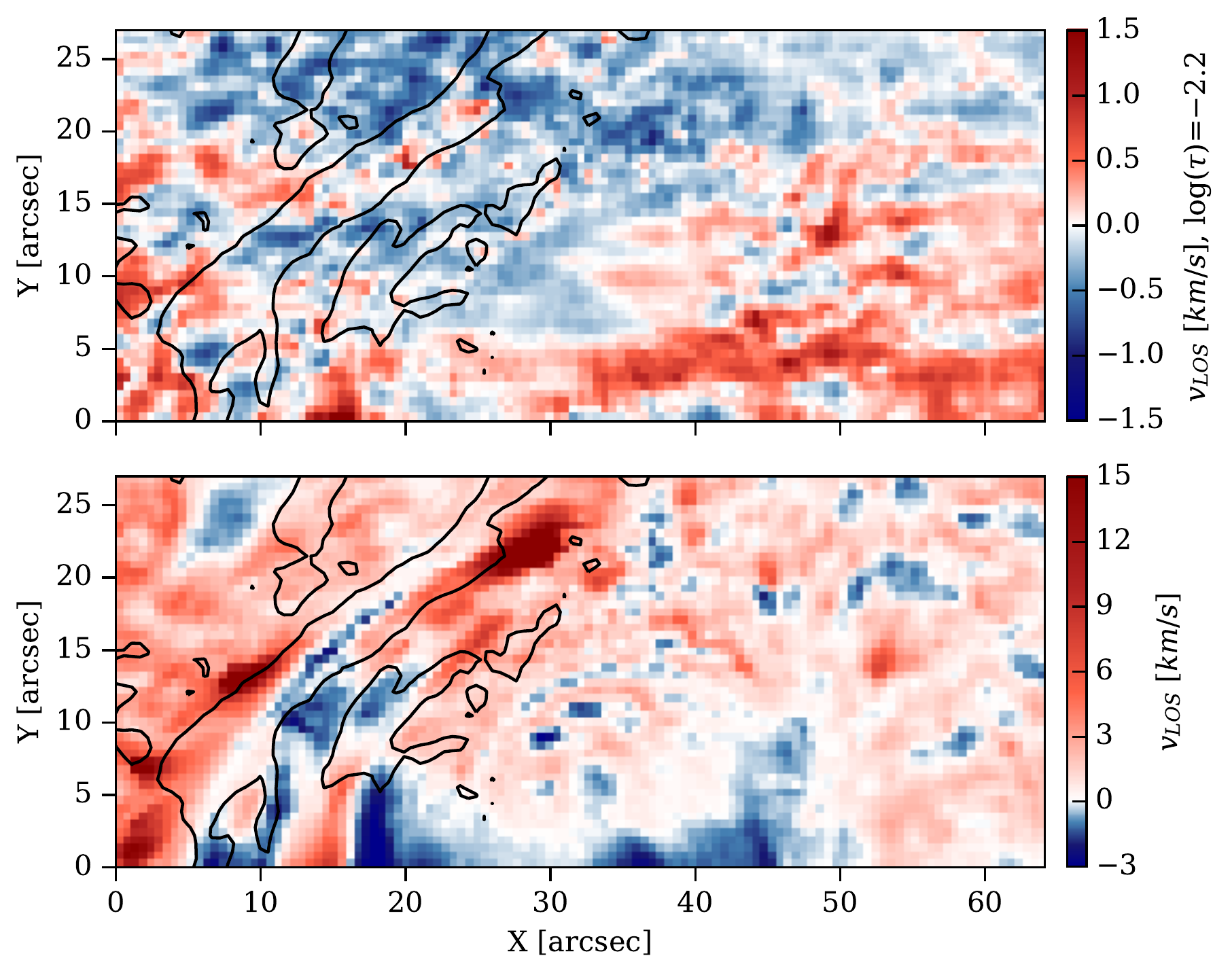}
\caption{LOS velocity from the inversion of the silicon line (upper panel) and the helium triplet (lower panel). Contours at 0.7$I_c$ are drawn to show the filament location. Negative values (blue) indicate motions towards the observer.}
\label{fig:vlos}
\end{figure}

Line-of-sight velocities obtained from the inversions in both atmospheric layers are shown in Fig.~\ref{fig:vlos}. The wavelength calibration was done using the photospheric spectral lines and comparing the intensity of an average quiet-Sun region in our data with the FTS atlas \citep{neckel1984}. In the photosphere, the umbra is essentially at rest with respect to the velocity of the quiet Sun. At the upper photospheric level ($\log(\tau)=-2.2$) the velocity in the FOV is between 1.5\,\kms\ and --1.5\,\kms.

In the chromosphere, weak upflows are found along the PIL with velocities around 3\,\kms. We identify this flow structure as such typical of slowly rising emerging loop-like structures \cite[e.g.,][]{solanki2003,xu2010}, suggesting that this is an emerging $\Omega$ loop, where cool material rises to chromospheric heights and then drains towards the footpoints of the loops. {That could be identified with the ends of the filament where downflows are detected}. The only difference between these arch filaments and our active region filament is that the filament material is aligned with the PIL, while their archs are mainly perpendicular. This scenario could be compatible with an emergent flux rope from the active region, presenting similar rising speeds to those of \cite{Kuckein2012V}.

The downflows are found at both sides of the filament with velocities larger than 20\kms. These pixels also show several (at least two) well separated components in Stokes $I$ and Stokes $V$. Pixels close to them do not show a clear two-component profile, but an extended wing towards red wavelengths. No clear indication of more than two components clearly separated in velocity is found, in contrast to \cite{sasso2011}.

Although a clear relation between the photosphere and chromosphere is hard to establish with our observations, it is tantalizing to associate the two small redshifted patches around (25\arcsec, 20\arcsec) in the photosphere with the chromospheric downflows, produced by material falling to the photosphere. These two patches in the photospheric map are localized to the left of the downflows because of the geometrical projection.

\subsection{Doppler width and optical depth}\label{sec:termo}
\begin{figure}[!ht]
\centering
\includegraphics[width=1.0\linewidth]{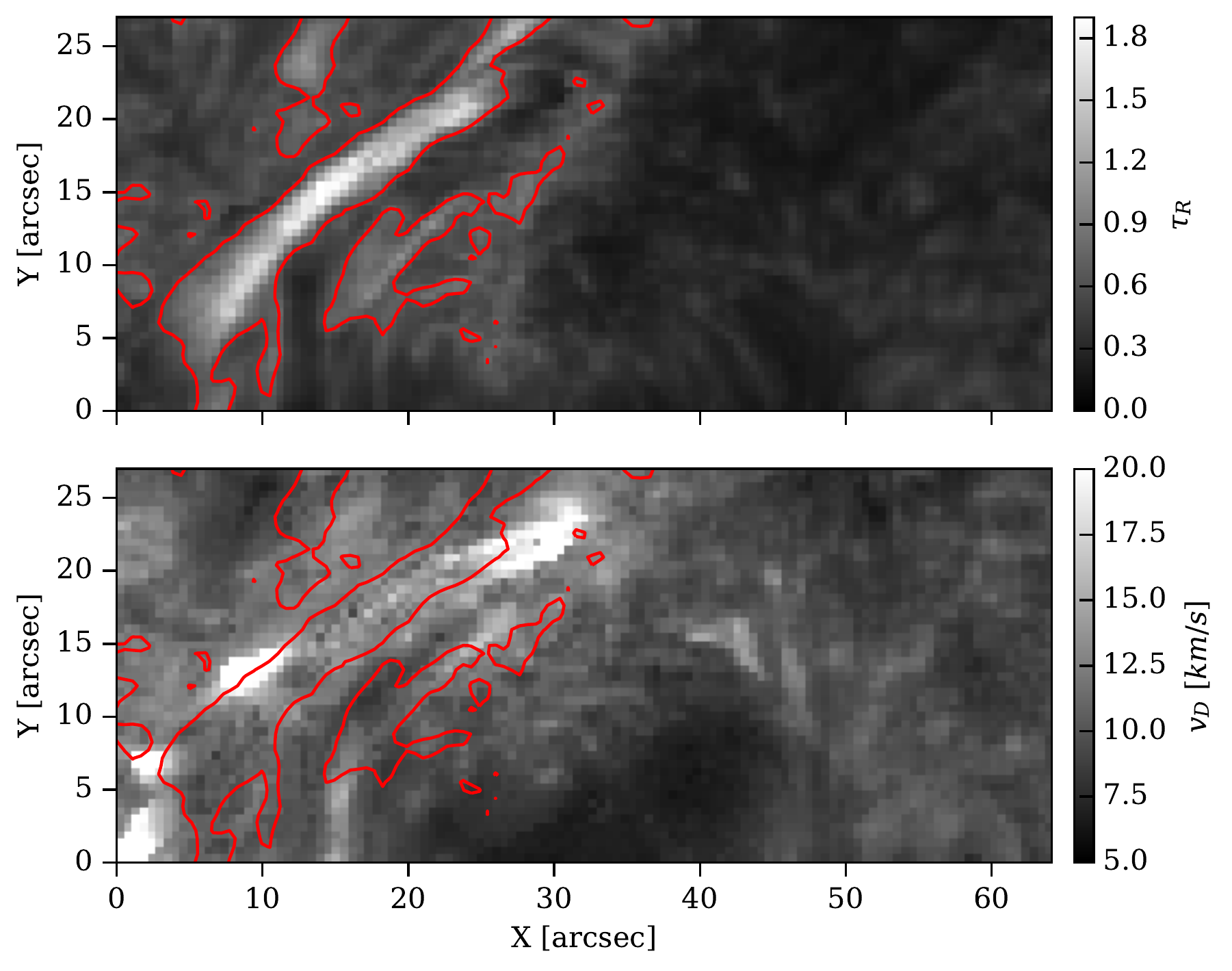}
\caption{Optical depth measured in the red component of the \ion{He}{i} 10830\,\AA\ (top panel). Doppler width as a result of fitting a single component (bottom panel). Larger values of Doppler width are obtained in the areas where more than one component appears in the Stokes profiles.}
\label{fig:tau}
\end{figure}

Figure~\ref{fig:tau} shows the optical depth $\tau_R$ and the Doppler 
width $\Delta v_D$ resulting from the inversion of the helium profiles. The mean optical depth in the filament is clearly above 0.7. The maximum value of the optical depth of the filament calculated in this scan is 1.9, but it easily goes up to 2.3 for other scans in which it is more opaque. Values above $\tau_R=1$ can be potentially affected by radiative transfer effects \citep{TB2007}, not taken into account in the version of \Hazel\ at the time of writing. These effects could potentially impact the inference of the magnetic field and its topology.

The Doppler width in the filament varies between 10\kms\ and 14\kms. Larger values are concentrated in the downflows and they are artificially produced by \Hazel\ misfits when  trying to fit a pixel with two line profiles clearly separated with a single atmospheric component model. Although the Doppler width of the line can also be  used as a tool to extract an upper limit to the temperature, we do not consider it to be a good idea. Many effects that produce unresolved velocities (like the fine structure of the filament with small-scale fibrils with potentially different plasma flows) also broaden the profiles and we lack the means to calibrate their importance.

\section{Bayesian inference}
\label{sec:bayesian}

It is clear from Secs.~\ref{sec:Blos10830}-\ref{sec:ambiguities}  that the inference of the magnetic field vector in the chromosphere requires a careful identification of all ambiguous solutions. Additionally, it is of paramount importance to capture all the potential degeneracies among the physical parameters and return reliable error bars. For this reason, we carry out a Bayesian analysis to fully exploit the observations.

The Bayesian formalism has been applied before in solar spectro-polarimetry, for instance analyzing the magnetic field in the quiet regions of the solar photosphere \citep{asensio2007,asensio2009}, for model comparison in spectropolarimetric inversions \citep{asensio2012}, or estimating the magnetic field strength from magnetograms \citep{asensio2015}. We present here the fundamental ideas of the formalism, although much more detailed information and instructive examples can be found in the mentioned studies (and references therein) and our Appendix \ref{sec:appendix}.

\subsection{Multimodal inference}
\label{sec:multimodal}

In the Bayesian framework, the most plausible model is the one that maximizes the posterior distribution. Given that several ambiguous solutions exist, the posterior probability distribution turns out to be multimodal. Each mode has its specific model parameter uncertainties, therefore, sampling the posterior distribution turns out to be difficult. The importance of every mode located in the posterior is quantified by its evidence $p(D)$\footnote{To compare the ''local'' evidence of different modes, this value is not integrated over the totality of the parameter space but on a region surrounding the mode. As the likelihood decreases away from the mode, the local evidence is mainly controlled by a region very close to the location of each mode.}. For two modes having the same maximum value of the posterior distribution (or, equivalently, the same likelihood when using flat priors), the evidence is larger the larger the volume of the space of parameters compatible with the observations. In other words, given equally good fits, the evidence is larger for those solutions where the parameters can vary more without compromising the fit. Thus, the evidence may be seen as a quantitative implementation of Occam's Razor. However, as we show later, all the solutions occupy a similar volume in the parameter space with similar likelihoods, the main difference being its location in the parameter space. Therefore, from all available modes explored during the posterior sampling, we choose the set of modes for which the difference between its log-evidence and that of the maximum is smaller than 6, following the Jeffrey's scale \citep{Jeffreys61}. We empirically checked that this threshold is ideal to distinguish all compatible solutions and discard any spurious mode obtained during the sampling.

When the posterior is as corrugated and multimodal as the ones we are dealing with, standard Markov-Chain Monte Carlo (MCMC) techniques can present problems and, for example, the widely-used Metropolis-Hastings algorithm \citep{metropolis1953}, could get  trapped in local minima. To explore the parameter space MCMC standard methods generate new points "close" to the previous one whose probability of acceptance depends on the likelihood ratio between the new and previous point. In this way, if the new point produces a larger likelihood it is accepted, while if it is worse it is accepted with a small probability depending on the ratio of likelihoods. On the other hand, the Nested Sampling method is a Monte Carlo technique in which, after drawing the points (known usually as live or active points in this method) from the prior, they are sorted by their likelihood. The smaller one is removed and replaced by a new point with higher likelihood. In this way, we are continuously sampling layers with larger likelihood independently of the location of the previous points.

For this reason, we have used the \MultiNest\ algorithm \citep{feroz2009}, a Bayesian inference tool tailored to explore multimodal posteriors and estimate the evidence of each mode. \MultiNest\ implements a nested-sampling Monte Carlo algorithm \citep{skilling2004} that is designed to calculate efficiently the evidence (Eq.~\ref{eq:evidence}) and simultaneously sample the posterior distribution when the later is multimodal and/or has strong degeneracies. During the sampling \MultiNest\ infers the appropriate number of clusters from the set of points to separate each mode. Other details about this procedure are described in the original article \citep{feroz2009}. By merging together \MultiNest\ and \Hazel, we have performed, for the first time, a Bayesian inference for observations in the \ion{He}{i} multiplet\footnote{We have used \MultiNest\ through the Python wrapper \PyMultiNest\ \citep{pymultinest2014} available on \url{http://github.com/JohannesBuchner}.}.

One of the few free parameters of \MultiNest\ is the number of live points $N_\mathrm{live}$, which is directly proportional to the final precision on the estimation of the evidence. A larger number provides a more accurate sampling but increases the computing time.  After exhaustive tests, we have verified that $N_\mathrm{live}=1000$ gives enough precision for our purposes. Concerning the computing time, the inference for each pixel takes of the order of 20 minutes, although it strongly depends on the number of free physical parameters.

\subsection{Bayesian analysis of magnetic ambiguities}\label{sec:multiambiguo}
We analyze in detail one set of profiles that are representative of the observations in terms of the number of compatible solutions. We study the ambiguities in the $(B,\theta_B,\phi_B)$ subspace. The rest of thermodynamic parameters are estimated by a classical least squares fit of Stokes $I$. We also infer the uncertainties $\sigma_{QU}$ and $\sigma_V$ from the observations by including them in the Bayesian analysis. This helps us to absorb any Gaussian-distributed systematic effect in the observations, in these uncertainties.

We have used uninformative flat priors for all the parameters. The priors span the ranges ($0,2000$)~G, ($0,\pi$), and ($-\pi,\pi$) for the modulus, inclination, and azimuth of the magnetic field, respectively. We impose uniform priors in the range $(-7,-8.5)$ for the natural logarithm of the uncertainties, which correspond to a standard deviation in the range $(2,9) \times 10^{-4}$ in units of the continuum intensity\footnote{The Bayesian framework does not only allow us to calculate the uncertainties of the parameters of the model but also infer the uncertainties of the observations given the model. This parameter absorbs not only the noise of the observations (that sometimes is difficult to calculate if we do not have enough continuum wavelength points) but also the inherent sub-optimality of the model itself  because it cannot reproduce some spectral features.}. This range is sufficient to correctly capture the posterior of the uncertainties and narrow enough to reduce the computing time with respect to having a wider range.

In order to gain a better insight about the number of ambiguous solutions and their  physical origin, we have found useful to do two inferences: one using only Stokes $Q$ and $U$, and another one using only Stokes $V$. The solution compatible with the full Stokes vector lies in the intersection of the posterior distributions of both inferences. In the following subsections we will study individual cases where four, three or one solution is compatible with the observed Stokes profiles.

After the inference and before the analysis, it is always a good practice to carry out posterior predictive checks using \Hazel\ to synthesize the Stokes profiles with the parameters inferred from the posterior. An example of this has been presented in Fig.~\ref{fig:zoology}, where we have drawn the Stokes profiles of two pixels with their maximum-a-posteriori best-fit and a green shadowed region indicating the range of possible solutions obtained from the posterior distribution.

\subsubsection*{Four solutions}
\begin{figure}[!ht]
\centering

\includegraphics[width=\linewidth]{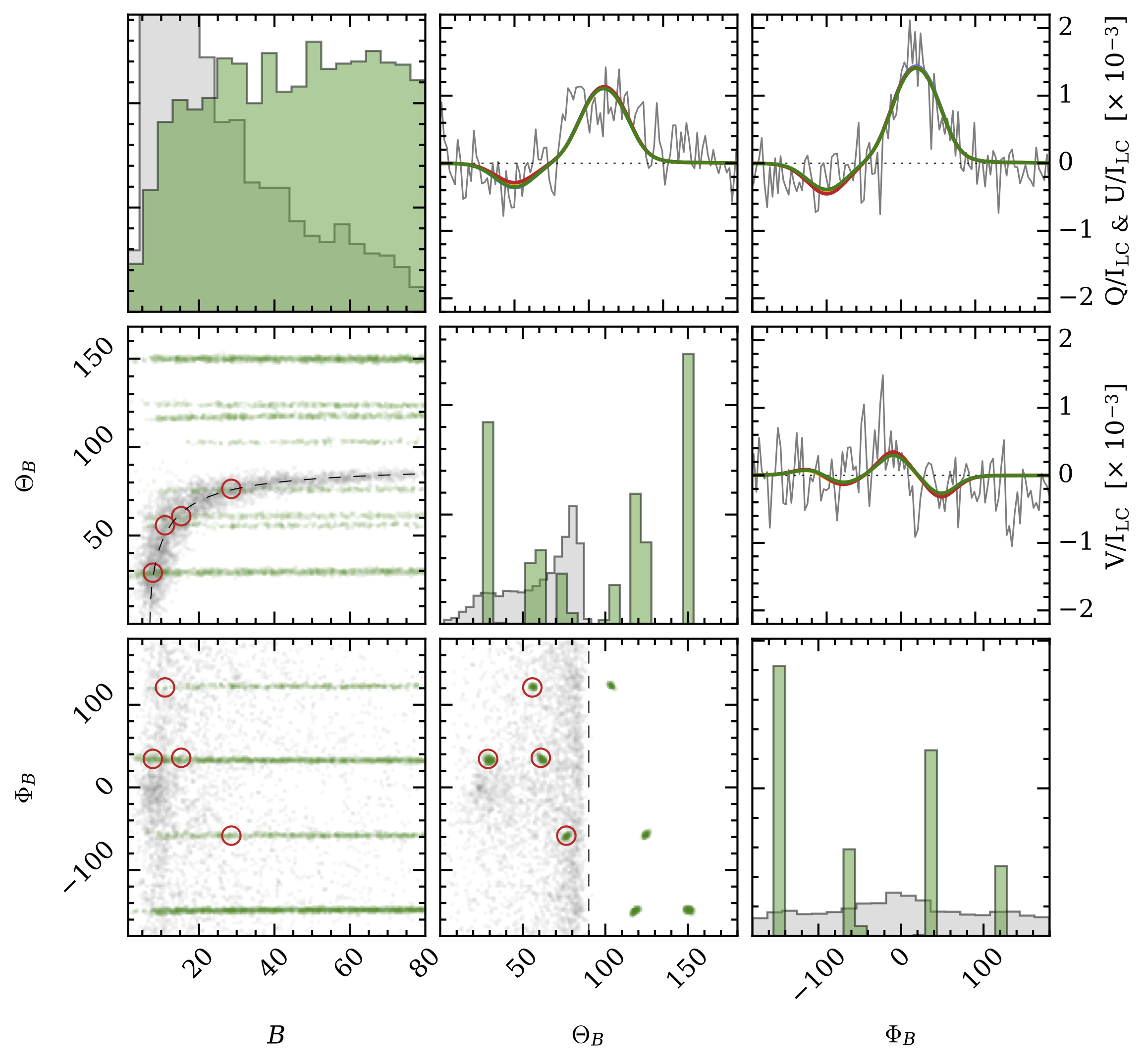}
\caption{Joint (below the diagonal) and marginal (in the diagonal) posterior distributions for the magnetic field strength, inclination, and azimuth in the LOS frame. The green samples and histograms correspond to the inference using only Stokes $Q$ and $U$, while the grey ones using only Stokes $V$. The analysis has been carried out for a pixel with four compatible solutions, drawn in different colors in the upper diagonal panels. The solutions compatible with the full Stokes profiles are indicated with red circles.}
\label{fig:4sol}
\end{figure}

The pixel at (24\arcsec, 25\arcsec) is a good example of a posterior with four modes. Figure~\ref{fig:4sol} displays the joint (below the diagonal) and marginal (in the diagonal) posterior distributions for the magnetic field strength and, inclination and azimuth in the LOS frame. The green samples and histograms correspond to the inference using only Stokes $Q$ and $U$, while the grey ones are obtained using only Stokes $V$. The presence of several modes is clearly seen in the joint posteriors. We have marked with red circles the solutions compatible with the full Stokes vector. The panels above the diagonal show the synthetic Stokes profiles for all the modes, demonstrating that they produce indistinguishable fits to the data.

The joint posterior distribution of the magnetic field strength and inclination in the LOS frame, $p(B,\Theta_B|D)$, when only Stokes $V$ is used presents a clear degeneracy, showing a hyperbolic-shaped posterior distribution typical of the Zeeman weak field regime, where only the product $B \cos\Theta_B$ can be reliably estimated. In order to facilitate the recognition of the degeneracy, we overplot the dashed line $B \cos\Theta_B=c$, where $c$ is a constant.

This pixel is in the saturation regime of the Hanle effect. In this regime, the Hanle effect is not sensitive to the magnetic field strength, being only sensitive to the orientation of the field. In this case, we have the following potential ambiguities: $\Phi_B$, $\Phi_B-\pi/2$,$\Phi_B+\pi/2$,$\Phi_B+\pi$. The green posterior (especially clear in the panel showing the joint posterior $p(B,\Phi_B|D)$) shows the four potential ambiguities in $\Phi_B$ as four horizontal lines.

In the joint posterior $p(\Theta_B,\Phi_B|D)$, we see how each degenerate value of the azimuth is compatible with two potential values for the inclination, generating a maximum of 8 possible ambiguous solutions. All the solutions occupy a similar volume in the parameter space, as it is seen from the green sampling, so the evidence is mainly controlled by the likelihood under our assumption of equal priors. Half of the points are discarded when the polarity of the field is taken into account in the inference through Stokes $V$. The sign of Stokes $V$ is such that only inclinations in the range $(0^\circ,90^\circ)$ are favoured by the observations. Therefore, we end up with the final four ambiguous solutions, marked with red circles. It is important to remember that some of the ambiguous solutions can have the same azimuth in the LOS and are only distinguished by their inclination.

For clarity, we only show the posterior distribution for the magnetic field strength up to 80\,G. However, the green posterior goes down to zero as we increase the magnetic field because the shape of Stokes $Q$ and $U$ does not fit the observations anymore. We note that the Hanle saturation regime starts at $\sim$10\,G and above this we are not sensitive to the strength of the magnetic field anymore and the posteriors are flat.

\subsubsection*{Three solutions}
A pixel representative of a posterior with three modes is the one at (4\arcsec, 5\arcsec). This pixel has three compatible solutions: two with the same LOS azimuth and one separated by $\pi/2$ to the other pair. 

\begin{figure}[!ht]
\centering
\includegraphics[width=\linewidth]{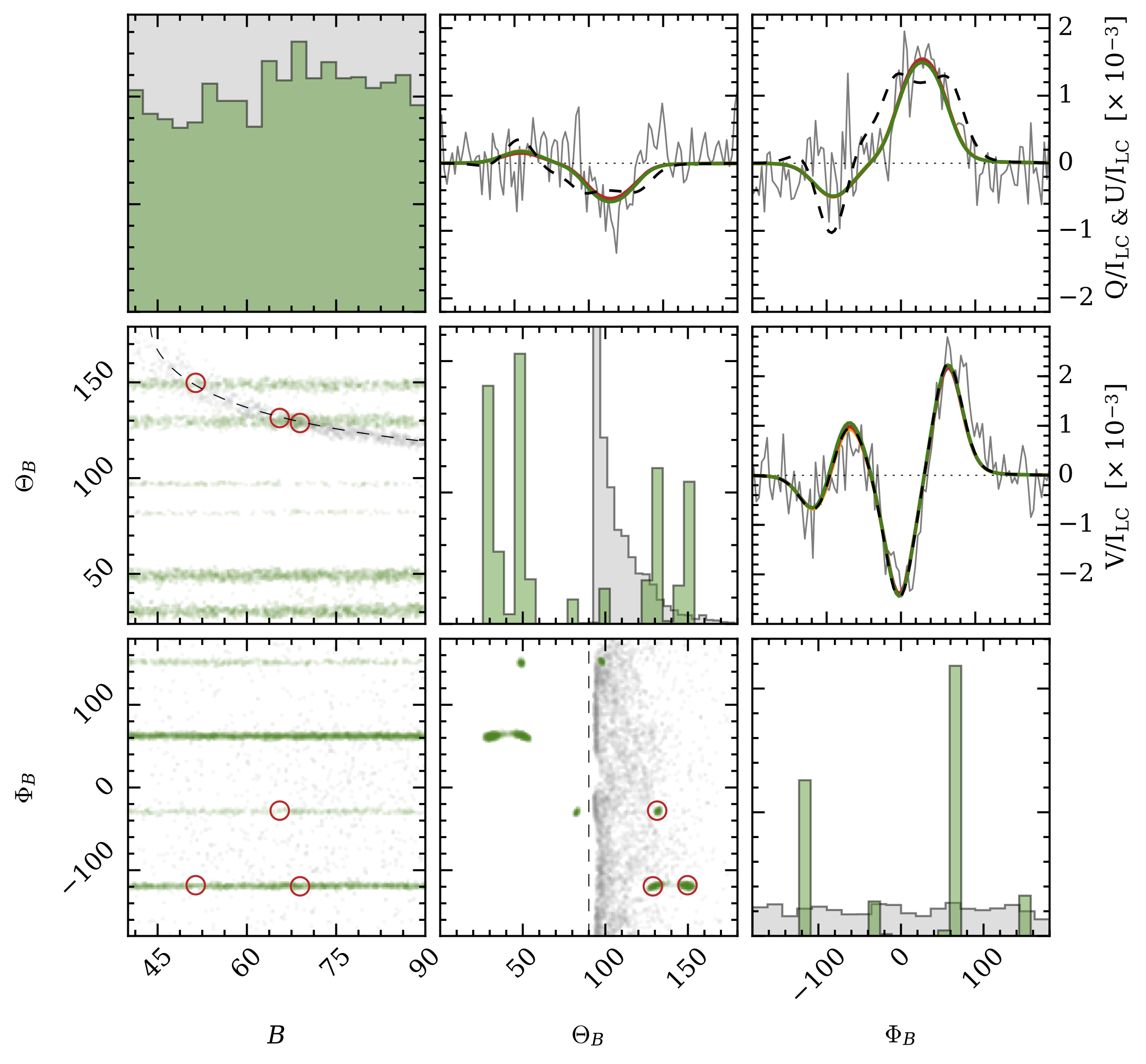}
\caption{
Same as Fig.~\ref{fig:4sol} but for a pixel with three compatible solutions. The dashed Stokes profiles come from the rejected solution with the same polarity. }
\label{fig:3sol}
\end{figure}

Figure~\ref{fig:3sol} displays the same set of quantities as in the previous example, but for this new pixel. In principle this pixel is in the saturation regime of the Hanle effect, but in this case only three of the four solutions (see the joint posterior $p(\Theta_B,\Phi_B|D)$) pass the Bayesian evidence test. The rejected point is very close to $\Theta_B=90^\circ$, so it is associated with a field almost perpendicular to the LOS. The ensuing magnetic field strength needed to produce the observed Stokes $V$ signal with this orientation of the field is close to 350\,G. This field is so strong that this solution is discarded because Stokes $Q$ and $U$ starts to show some Zeeman features that are not compatible with the observed profiles. This is clearly seen in the panel for Stokes $Q$, $U$, and $V$. The Stokes profiles of all three compatible solutions are displayed in the upper diagonal panels in solid lines, while the rejected solution is shown with the dashed line.

\subsubsection*{One solution}
\begin{figure}[!ht]
\centering
\includegraphics[width=\linewidth]{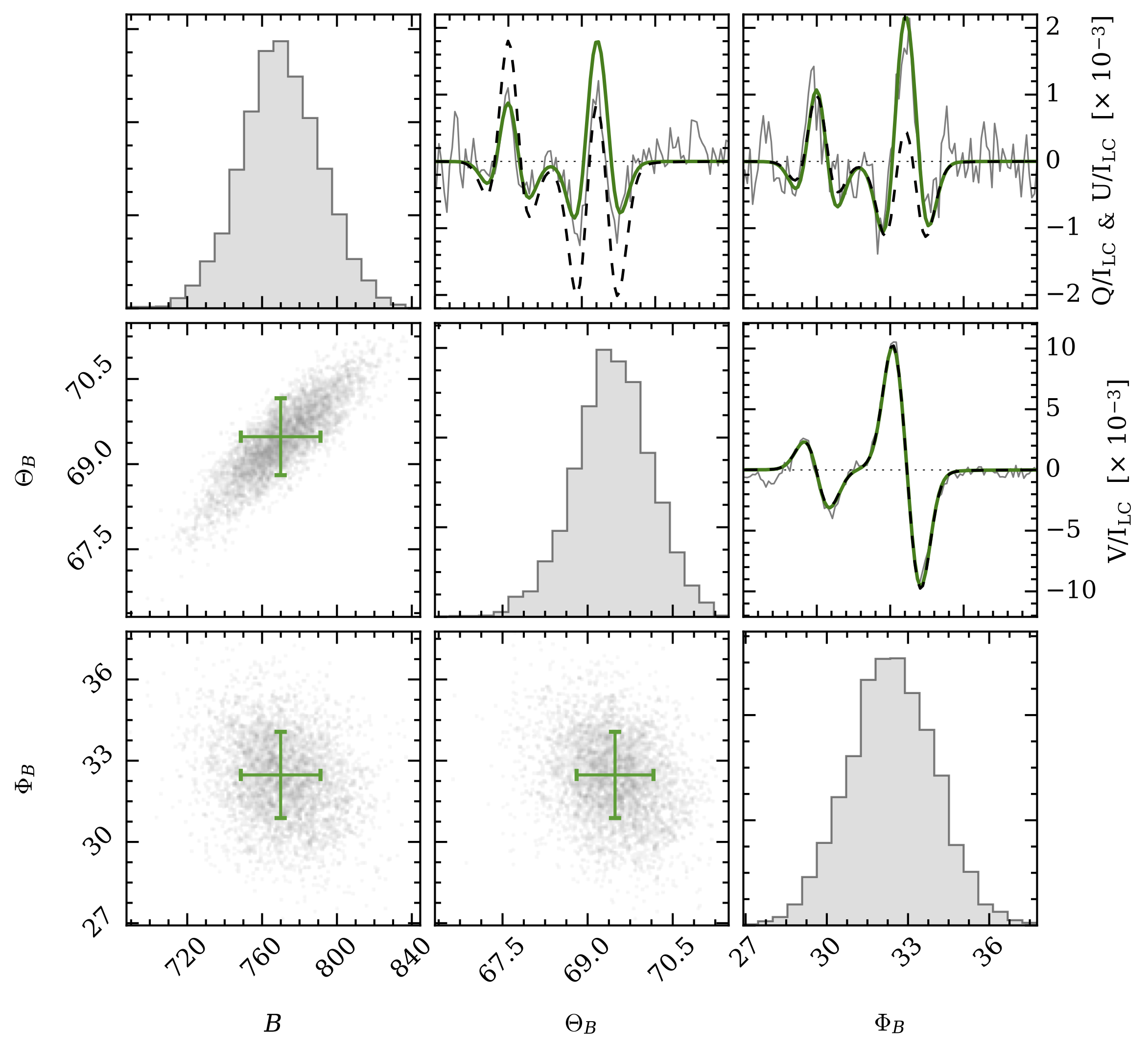}
\caption{
Same as Fig.~\ref{fig:4sol} but for a pixel with only one possible solution and using Stokes $Q$, $U$, and $V$ for the inference. Observed Stokes profiles (gray), synthetic profiles from the solution (green), and the rejected solution $\Theta_B+\pi$ (dashed) are shown in the upper diagonal panels.}
\label{fig:1sol}
\end{figure}

One of the most surprising conclusions of this work is the fact that it is possible to find pixels in which only one solution is compatible with the observations. A good example of this is the pixel at (38\arcsec, 8\arcsec). The joint and marginal posterior distributions of this pixel are displayed in Fig.~\ref{fig:1sol}. The profiles are clearly close to the pure Zeeman regime for Stokes $Q$ and $U$, with the typical three-lobed profiles, but some contribution from scattering polarization still  exists. In fact, the two effects act together and the $180^\circ$ ambiguity in the azimuth disappears \citep{schad2013}. The reason is that the Hanle contribution in the core of the Stokes $Q$ and $U$ has opposite sign in both solutions, which produces a misfit in one of them. The Zeeman contribution to Stokes profiles depends on the inclination and azimuth of the magnetic field with respect to the LOS, while for the Hanle contribution, it also depends on the geometry in the local reference frame. For our geometry ($\mu=0.92$) the two solutions with a difference in the azimuth of 180$^\circ$ lead to a different geometry in the local frame and only one of them is compatible with the Hanle signals. Therefore, the observing geometry plays a key role, allowing us in some cases to reduce the number of compatible solutions to a single one.

This ``self-disambiguation'' is only possible in the intermediate regime between the Hanle and the Zeeman effect ($\sim$400--800\,G), when the Hanle effect can generate slightly different profiles for the two solutions. The dashed lines in the Stokes profiles panels refer to the rejected solution ($\Phi_B+\pi$), being clearly unable to reproduce the profiles.

\subsection{Height of the filament}\label{sec:height}

When the observed structure is off the limb, imaging techniques can be used to estimate its height \citep{marian2015,schad2016}. On the contrary, for on-disk observations it is much more complicated since no straightforward technique for estimating the height is available. Using Bayesian inference we explore the feasibility of the inference of the height of the filament directly from the observations, as the larger the height, the larger the anisotropy, which induces a stronger zero-field scattering polarization signal. 

Some authors have tried to retrieve the height including it as an inversion parameter \citep{xu2010,merenda2011}, concluding that it can be inferred with high accuracy, solving possible debates in the interpretation of the topology \citep{Judge2009}. Additional arguments based on 3D geometrical reconstructions \citep{xu2010,solanki2003} have been used to calculate this height. However, it was noted by \cite{asensio2008} that the height cannot be easily inferred from the observations unless some information about the topology of the field is fixed a priori (for instance, the inclination). Here, we demonstrate that this is not only the case, but the inference is even more difficult because both the magnetic field and the optical depth are degenerate with the height.

\begin{figure*}[!ht]
\centering
\includegraphics[width=\linewidth]{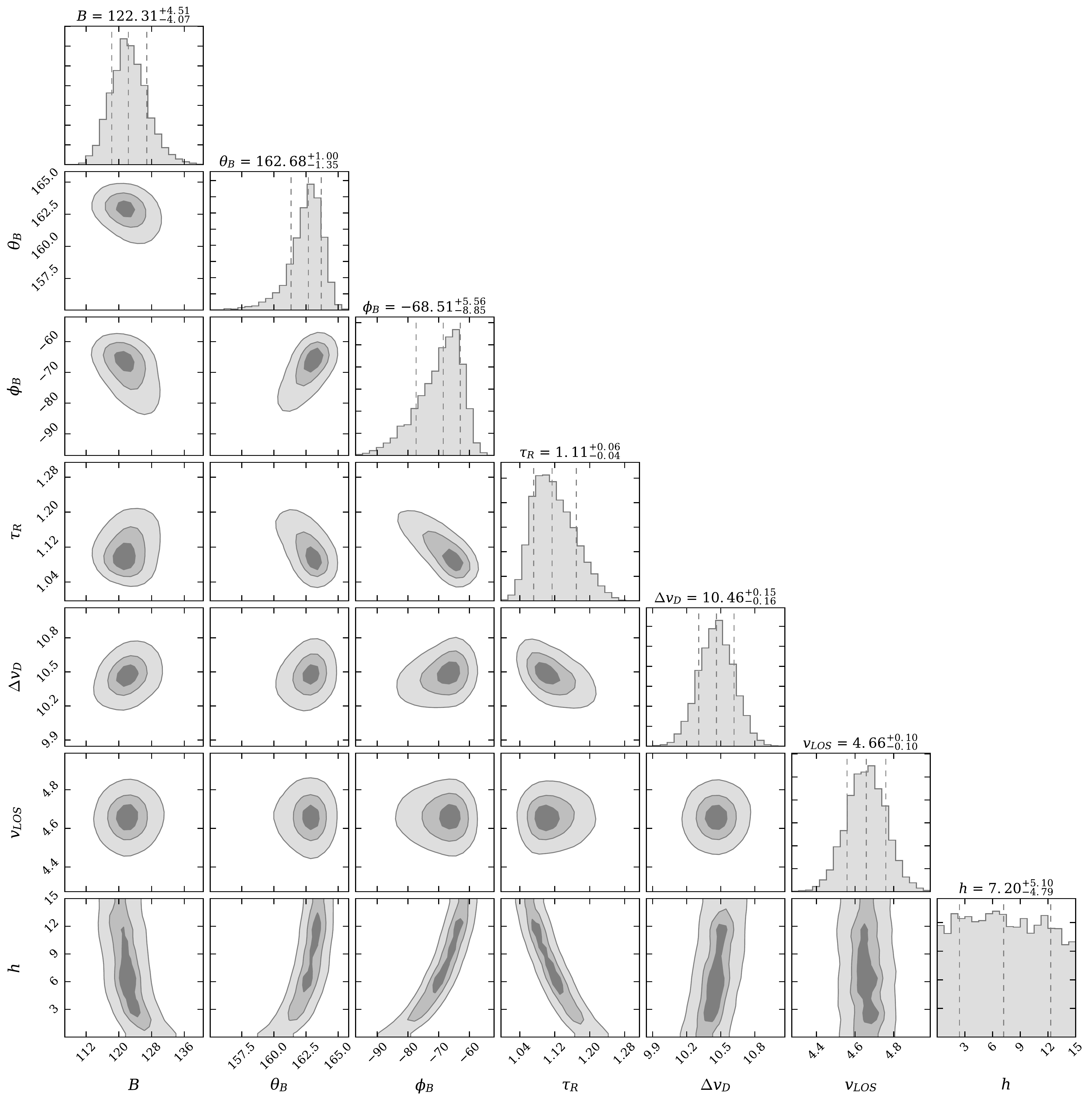}
\caption{
Histograms and correlations of the posterior distributions of each physical parameter. The label on top of each column provides the median and the uncertainty defined by the percentiles 16 and 84 (equivalent to the standard 1$\sigma$ uncertainty in the Gaussian case). Also the contours are shown at 0.5, 1, and 2 sigma.}
\label{fig:mtutti}
\end{figure*}

To analyze the reliability of inferring the height from the Stokes parameters, we choose the pixel at (22\arcsec, 22\arcsec) which lies inside the filament. We include all 10 free variables: $B$, $\theta_B$, $\phi_B$, $\tau_R$, $\Delta v_D$, $v_{LOS}$, $h$, $\sigma_I$, $\sigma_{QU}$, and $\sigma_V$ in our Bayesian analysis. Given that the number of free parameters is large and to decrease the computing time, we use more restrictive priors for the inclination than before, trying to capture only one possible ambiguous solution. The prior for the height is flat and limited to be in the range (0\arcsec, 15\arcsec). We carried out some tests to check that isolating one mode of the magnetic field does not affect our conclusion. As before, we add the uncertainty of the Stokes parameters as variables and impose a suitable Jeffreys' prior. Sampling the unimodal posterior requires more than 100 minutes on a single CPU (3.4GHz Intel Core i7).

The ensuing posterior distributions (joint and marginal) are displayed in 
Fig.~\ref{fig:mtutti}\footnote{We have removed from the figure the distributions of $\sigma_I$, $\sigma_{QU}$, and $\sigma_V$ to improve its readability.}. The posterior distribution seems to be quite localized for all variables, except for the height. Its marginal posterior is practically flat, almost indistinguishable from the assumed prior. Concerning the joint posteriors, it displays a very strong correlation with many of the remaining parameters. For instance, changing the height by 15\,\arcsec\ can be compensated easily by barely modifying the optical depth from 1.2 to 1.04. The same happens with the inclination of the field, where a small change of 4$^\circ$ compensates the change in height. Finally, changes of only 10\,G in the magnetic field strength can also compensate for the same change in height, in agreement with \cite{schad2013}.  {Therefore, this analysis shows that while height cannot be inferred accurately because its effect can be easily compensated for with other parameters, the rest of parameters are obtained with very small uncertainty.}

\section{Summary and conclusions}

In this article we have shown a first analysis of ground-based spectropolarimetric observations acquired with GRIS (GREGOR, at Observatorio del Teide) of the \ion{He}{i} 10830 triplet in an active region filament. The active region was located in a flaring area. Although the profiles do not show any peculiar signature, the fast evolution of the filament (with,  e.g., the absorption strongly changing in consecutive scans, $\sim$15 min) is a sign of the activity of the area.

The observed polarization signals in the filament show single-lobed linear polarization profiles, which is indicative of optical pumping by an anisotropic illumination and, potentially, the action of the Hanle effect. These linear polarization signatures were close to the noise level ($5\times10^{-4}I_c$). We have been able to measure them as a result of our high polarimetric sensitivity and a noise reduction process. Longer integration times than the ones used here are necessary to increase the signal-to-noise ratio and study the weak polarization signals present in these chromospheric structures. The low signal-to-noise ratio of the observations from other studies, such as \cite{xu2012}, might be the reason why they could not detect the linear polarization signals due to scattering, hence neglecting atomic polarization in their analysis.

We have studied the observations with \Hazel, a state-of-the-art inversion code, which includes the Hanle and Zeeman effects, to retrieve the physical properties from the polarimetric signals. Using a one component model, as it is customary in other studies, magnetic field strength values around 50--150\,G have been found. 

We have also applied the Bayesian framework using \Hazel\ and \MultiNest\ to explore the solutions in the parameter space and compute posterior distributions. This Bayesian inversion cannot compete in speed with the standard inversion algorithms but it can be used to investigate in detail the accuracy of the inversions, the sensitivity of the parameters to the noise, and to give confidence intervals to all the parameters. Using this technique, we have studied how many solutions are compatible with the analyzed profiles depending on the geometry of scattering and the magnetic field regime. In addition, we have applied it to the inference of the height, concluding that this parameter cannot be constrained, at least at our noise level.

Some results clearly point out that a more complex inversion than a single component model had to be done. First, the Stokes $V$ map of \ion{He}{i} does not show any clear signature of the presence of the filament. Second, the inferred azimuth map follows the same pattern as Stokes $V$, as if the polarity of Stokes $V$ were conditioning the inference. This indication suggests that the Stokes $V$ could be dominated by magnetic field of the underlying active region, and not, from the filament itself \citep{diaz2016}. Then, a more complex model is needed to explain the observations and avoid misinterpretations due to oversimplified models \citep{milic2016}. A deeper study of these features is done in the second part of this series, studying in detail all the evidences of the limitations of a single component model.

\begin{acknowledgements}

We would like to thank the anonymous referee for its comments and suggestions. The authors wish to thank Manolo Collados for useful advice on the presented work. We specially thank Andreas Lagg for observing the active region filament and providing this valuable data for us.

The 1.5-meter GREGOR  solar telescope was built by a German consortium under the leadership of the Kiepenheuer Institut f\"ur Sonnenphysik in Freiburg with the Leibniz Institut f\"ur Astrophysik Potsdam, the Institut f\"ur Astrophysik G\"ottingen, and the Max-Planck Institut f\"ur Sonnensystemforschung in G\"ottingen as partners, and with contributions by the Instituto de Astrof\'sica de Canarias and the Astronomical Institute of the Academy of Sciences of the Czech Republic. The GRIS instrument was developed thanks to the support by the Spanish Ministry of Economy and Competitiveness through the project AYA2010-18029 (Solar Magnetism and Astrophysical Spectropolarimetry).

Financial support by the Spanish Ministry of Economy and Competitiveness 
through projects AYA2014-60476-P, AYA2014-60833-P and Consolider-Ingenio 2010 CSD2009-00038 are gratefully acknowledged. 

CJDB acknowledges Fundaci\'on La Caixa for the financial support received in the form of a PhD contract. 

This research has made use of NASA's Astrophysics Data System Bibliographic Services.

This paper made use of the IAC Supercomputing facility HTCondor (\url{http://research.cs.wisc.edu/htcondor/}), partly financed by the Ministry of Economy and Competitiveness with FEDER funds, code IACA13-3E-2493.

We acknowledge the community effort devoted to the development of the following open-source packages that were used in this work: \texttt{numpy} (\texttt{numpy.org}), \texttt{matplotlib} (\texttt{matplotlib.org}), \texttt{corner} \citep{corner}, and \texttt{SunPy} (\texttt{sunpy.org}).

\end{acknowledgements}

\bibliographystyle{mnras}

\begin{appendix}
\section{Bayesian framework}\label{sec:appendix}
Bayesian inference provides the most complete approach to parameter inference. The inference problem relies on the calculation of the posterior probability distribution function (PDF) $p(\thetabold|D)$ of a set of $M$ parameters $\thetabold$ (in this case the vector of physical quantities of the model) that is used to describe a given observation $D$. This posterior distribution describes our updated state of knowledge on the parameters after taking the data $D$ into account. Bayes' theorem allows us to calculate the posterior distribution for $\thetabold$:
\begin{equation}
 p(\thetabold|D) = \frac{p(\thetabold)p(D|\thetabold)}{p(D)},
 \label{eq:bayes}
\end{equation}
where $p(\thetabold)$ is the prior and represents our a priori knowledge in the model parameters. It is usual to have some initial information about the physical parameters. For instance, an estimation of the range of variation of the physical parameters, although sometimes it can be a very rough range (for instance, only positive values). This information is incorporated into a prior distribution. In the most simple case, we can assume that all the parameters are statistically independent, so that the prior distribution can be written as:

\begin{equation}
 p(\thetabold) = \prod_{i=0}^Mp(\theta_i),
 \label{eq:prior}
\end{equation}
where the $\thetabold$ are the parameters included in the model and $M$ is the number of such parameters. In the case that the only physical information available is the range of variation of the parameters we will use  uninformative priors with an uniform probability distribution inside the range and zero otherwise.

The second term of the Bayes' equation $p(D|\thetabold)$ is the so-called likelihood, that measures the probability that the data $D$ were obtained (measured) assuming given values for the model parameters $\thetabold$, i.e., gives information about how well the parameters predict the observed data.  In the case of the inversion of Stokes profiles, the data $D$ that we are facing consists on a set of four vectorial quantities, i.e., the wavelength dependence of the four Stokes parameters.

Under the assumption of uncorrelated Gaussian noise in the data\footnote{It is
true that some remaining correlation might exist among the measured Stokes parameters because
they are in fact obtained with a modulation process. Therefore cross-talk and other instrumental effects might "correlate" the noise in the four Stokes parameters.}, the likelihood is
given by:
\begin{equation}
  p(D|\thetabold) = \prod_{i=0}^3 \prod_{j=1}^N \frac{1}{\sigma_i\sqrt{2\pi}} \text{exp} \left[ -\frac{[\,S_i^\mathrm{obs}(\lambda_j) - S_i^\mathrm{syn}(\thetabold,\lambda_j)\,]^2} {{2}\sigma_i^2(\lambda_j)} \right],
 \label{eq:likelihood}
\end{equation}
where $S_i^\mathrm{obs}(\lambda_j)$ is the observed Stokes parameter $\Sbold=(S_0=I,S_1=Q,S_2=U,S_3=V)$ at wavelength $\lambda_j$, $S_i^\mathrm{syn}(\thetabold,\lambda_j)$ is the emergent synthetic Stokes parameter when the forward problem is solved in a given atmospheric model parameterized by the vector of parameters $\thetabold$, and $\sigma_i(\lambda_j)$ is the standard deviation of the noise for each Stokes parameter and wavelength point. Strictly speaking, the noise in the observed Stokes profiles follows a Poisson distribution because it comes mainly from photon noise. However, for consistency with other works, we choose the noise to be normally distributed, which will be a good approximation if the number of photons is high enough. Although we have focused on wavelength-independent noise, the formalism allows to straightforwardly accommodate wavelength-dependent noise in the likelihood. Therefore, if the information is available, it is possible to include other sources of uncertainty like data reduction residuals, cross-talk, fringes, etc.


Finally, the denominator $p(D)$ in Eq.~\ref{eq:bayes} is the Bayesian evidence or 
marginal posterior, which is a normalization factor that has turned out to be
important in our analysis, as shown in the main text. Since the posterior is a
probability distribution, its total integral (or sum, if the space is
discrete) must equal one. Therefore the evidence can be written as:
\begin{equation}
  p(D)= \int{p(D|\thetabold)\,p(\thetabold)\,d\thetabold}\,.
 \label{eq:evidence}
\end{equation}
For parameter inference, the information for a single parameter can be obtained 
from the marginal posterior, obtained from the posterior by integrating over the rest 
of parameters (marginalization):
\begin{equation}
  p(\theta_i|D)= \int{p(\thetabold|D)}\ d\theta_1...d\theta_{i-1}d\theta_{i+1}...d\theta_{M}\, .
 \label{eq:marginalization}
\end{equation}

The two–dimensional posterior distribution where we have marginalized the whole posterior distribution to all the parameters less two of them, i.e.:
\begin{equation}
  p(\theta_i,\theta_j|D)= \int{p(\thetabold|D)}\ d\theta_1...d\theta_{i-1}d\theta_{i+1}...d\theta_{j-1}d\theta_{j+1}...d\theta_{M}\,
 \label{eq:marginalization2}
\end{equation}
is commonly known as joint posterior.

Once the posterior distribution $p(\thetabold|D)$ is known, the position of the maxima (in our case the posterior is multimodal) gives the most probable combination of parameters that reproduce the observational data. Moreover, we can analyze in detail the confidence levels of the parameters, degeneracies, ambiguities, and the rest of problems that arise in typical inversion codes. Therefore, our objective is then to sample the posterior distribution and to find the combination of parameters that produce these maxima.




\end{appendix}

\end{document}